\documentclass[a4paper,10pt]{article}
\usepackage[utf8]{inputenc}
\usepackage{jheppubMod}
\usepackage{amsmath,amsfonts,amssymb,graphicx,MnSymbol,subfigure,multirow}
\usepackage{hyperref} 
\usepackage{tikz}
\usetikzlibrary{shapes.geometric, arrows}
\usepackage[toc,page]{appendix}
\usepackage{multirow,arydshln}

\usepackage[utf8]{inputenc}
\usepackage{amsmath,amsfonts,amssymb,graphicx,MnSymbol,subfigure,multirow}
\usepackage{hyperref} 

\title{ Semi-visible emerging jets }

\abstract{
    We propose a new class of dark-shower signatures in Standard Model extensions featuring Hidden Valleys or dark sectors coupled through an s-channel mediator. In this framework, unstable dark pions appear as long-lived particles (LLPs), with their lifetimes treated as free parameters. The resulting signatures, which we term semi-visible emerging jets (SVEJ), continuously interpolate between the established semi-visible and emerging jet regimes. We outline an analysis strategy optimized for dark pion lifetimes of order $\mathcal{O}(10)$ mm, and reinterpret existing LLP searches targeting lifetimes of $\mathcal{O}(100)$–$\mathcal{O}(1000)$ mm. Our proposed SVEJ search, exploiting the current ATLAS emerging-jet trigger, achieves sensitivity to cross sections of $\mathcal{O}(0.1)$ fb for lifetimes around $\mathcal{O}(10)$ mm. Finally, we advocate a more detailed study, including hadronisation uncertainties and detector-level effects.}

\author[a]{Juliana Carrasco,}
\author[b]{Suchita Kulkarni,}
\author[c]{Wei Liu,}
\author[b]{Joshua Lockyer,}
\author[a]{Jose Zurita,}
\affiliation[a]{\it Instituto de F\'{\i}sica Corpuscular, CSIC-Universitat de Val\`encia, Valencia, Spain}
\affiliation[b]{Institute of Physics, NAWI Graz, University of Graz, Universit\"atsplatz 5, A-8010 Graz, Austria}
\affiliation[c]{Department of Applied Physics and MIIT Key Laboratory of Semiconductor Microstructure and Quantum Sensing, Nanjing University of Science and Technology, Nanjing 210094, China}
\emailAdd{juliana.carrasco@ific.uv.es}
\emailAdd{suchita.kulkarni@uni-graz.at}
\emailAdd{wei.liu@njust.edu.cn}
\emailAdd{joshua.lockyer@uni-graz.at}
\emailAdd{jzurita@ific.uv.es}

\newcommand{\newc}{\newcommand}
\newcommand{\ld}{{\Lambda}}
\newcommand{\mpl}{{m_\pi/\ld}}
\newcommand{\nc}{{N_C}}                
\newcommand{\nf}{{N_F}}                
\newcommand{\fc}{{\nf/\nc}}                
\newcommand{\pid}{{\pi^0}}                
\newcommand{\mpid}{{m_{\pi^0}}}         
\newcommand{\mzp}{{m_{Z^\prime}}}
\newcommand{\ctp}{{c\tau_{\pid}}}

\newc{\UV}{{\mathchoice{}{}{\scriptscriptstyle}{}UV}}
\newc{\IR}{{\mathchoice{}{}{\scriptscriptstyle}{} IR}}

\begin{document}
\maketitle


\section{Introduction}


LHC Run-3 has provided an unprecedented opportunity to explore non-standard signatures that may have escaped detection in earlier runs. In parallel with the growing dataset, the search for new physics has become increasingly attentive to non-prompt, composite, and unconventional final states, scenarios where dark sectors interact weakly with the Standard Model (SM) yet leave subtle imprints in detector signatures. Among these, dark showers provide one of the most compelling frameworks. They arise naturally in models featuring hidden confining gauge dynamics, where a mediator, typically a heavy portal such as a dark photon or scalar, connects the Standard Model to a sector that hadronizes under its own non-Abelian force. Such constructions are commonly referred to as confining Hidden Valley (HV)~\cite{Strassler:2006im, Strassler:2006qa} or dark sector (DS) models and will be referred to as HV/DS scenarios in what follows.

HV/DS scenarios have strong theoretical motivations, which include the Twin Higgs, its variants~\cite{Chacko:2005pe,Craig:2015pha, Craig:2014aea, Craig:2016kue, Craig:2014roa}~\footnote{Recently there is also interest in the SM Higgs serving as a mediator in HV/DS context, see e.g.~\cite{Lu:2023gjk, Carrasco:2023loy,Cheng:2024aco}.}, dark matter~\cite{Cline:2021itd,Hochberg:2014kqa,Bernreuther:2023kcg,Pomper:2024otb,Kolesova:2025ghl,Kulkarni:2022bvh}, and baryogenesis~\cite{Craig:2010au,Katz:2016adq}. They also have interesting connections to gravitational-wave signatures~\cite{Schwaller:2015tja,Huang:2020crf,Pasechnik:2023hwv}, and inflationary paradigms~\cite{Li:2021fao, Cacciapaglia:2023kat}. These motivations make a compelling case to study them in light of our understanding of non-perturbative physics.

In addition to this, the HV/DS scenarios offer new signatures at colliders. In the last few years, both ATLAS and CMS have established a search program towards well-known HV/DS signatures. These signatures include the semi-visible jets~\cite{Cohen:2015toa,Cohen:2017pzm,Beauchesne:2017yhh} -- targeting the prompt decay of the bound states, emerging jets~\cite{Schwaller:2015gea,Renner:2018fhh} -- targeting long-lived bound states. Apart from jets, these scenarios may also produce more exotic final states such as soft-unclustered-energy-patterns~\cite{Harnik:2008ax,Polchinski:2002jw,Hatta:2008tx,Knapen:2016hky,Harnik:2008ax,Cesarotti:2020uod,Lu:2025cty}. Results from the first experimental searches for semi-visible jets~\cite{CMS:2021dzg,ATLAS:2023swa,ATLAS:2023kao,ATLAS:2025kuz}, emerging jets~\cite{CMS:2018bvr, CMS:2024rea,CMS:2024gxp,ATLAS:2025bsz,ATLAS:2025lfx}, displaced dimuons~\cite{CMS-PAS-EXO-24-008} and soft-unclustered-energy-patterns~\cite{CMS:2024nca} are also available. These searches and signatures are guided by phenomenological benchmarks rather than by models grounded in consistent hidden-sector dynamics. For a review on strongly-coupled theories see e.g.~\cite{Albouy:2022cin,Kribs:2016cew,Cacciapaglia:2020kgq,Cline:2021itd}. 

In the current classification scheme, the diverse signatures appear disconnected, even though they may arise from common underlying dynamics. For example, the emerging jets signature is often discussed in the context of t-channel mediators, while the semi-visible jets are thought to be characteristic of s-channel scenarios. In reality, this parameter space forms a continuum that becomes apparent when the signature space is guided by theoretically grounded top-down model constructions. Although top-down constructions are inherently more model dependent, they make it possible to systematically relate classes of models to characteristic collider signatures, revealing which theoretical features control specific observables. For an example of semi-visible and emerging jets appearing together for t-channel models see~\cite{Carmona:2024tkg} and in context of glueballs see~\cite{Curtin:2022tou}.

The top-down construction approach has another advantage. Unlike weakly interacting, non-confining dark sectors, a top-down approach to hidden valley or dark-shower phenomenology reveals theoretical features that phenomenological models cannot capture. The characteristic hierarchy of scales, the running of the coupling governed by ultraviolet parameters, and the resulting parton-shower and hadronization dynamics, each correlated with the underlying high-scale dynamics, are key examples of effects that only emerge in consistent confining constructions.

Motivated by these principles, in this work, we concentrate on a specific class of top-down HV/DS models. We consider an HV/DS sector consisting of an SU$(\nc)$ gauge group with $\nf$ flavours. This sector communicates with the SM via an s-channel HV/DS flavour conserving $Z^\prime$. Owing to the small coupling between the HV/DS and SM sectors, the HV/DS mesons are naturally long-lived. We argue that this class of models generically feature a new class of signatures we term as semi-visible emerging jets. 

\begin{figure}[h!]
    \centering
    \includegraphics[width=0.45\linewidth]{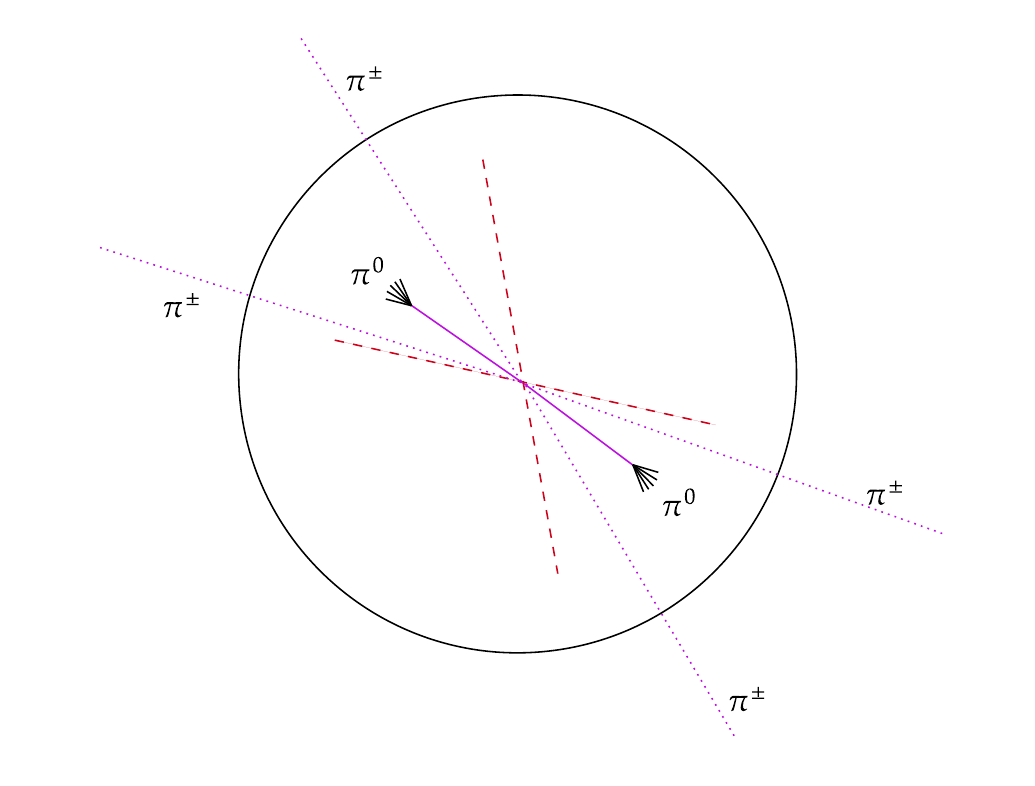}
    \includegraphics[width=0.45\linewidth]{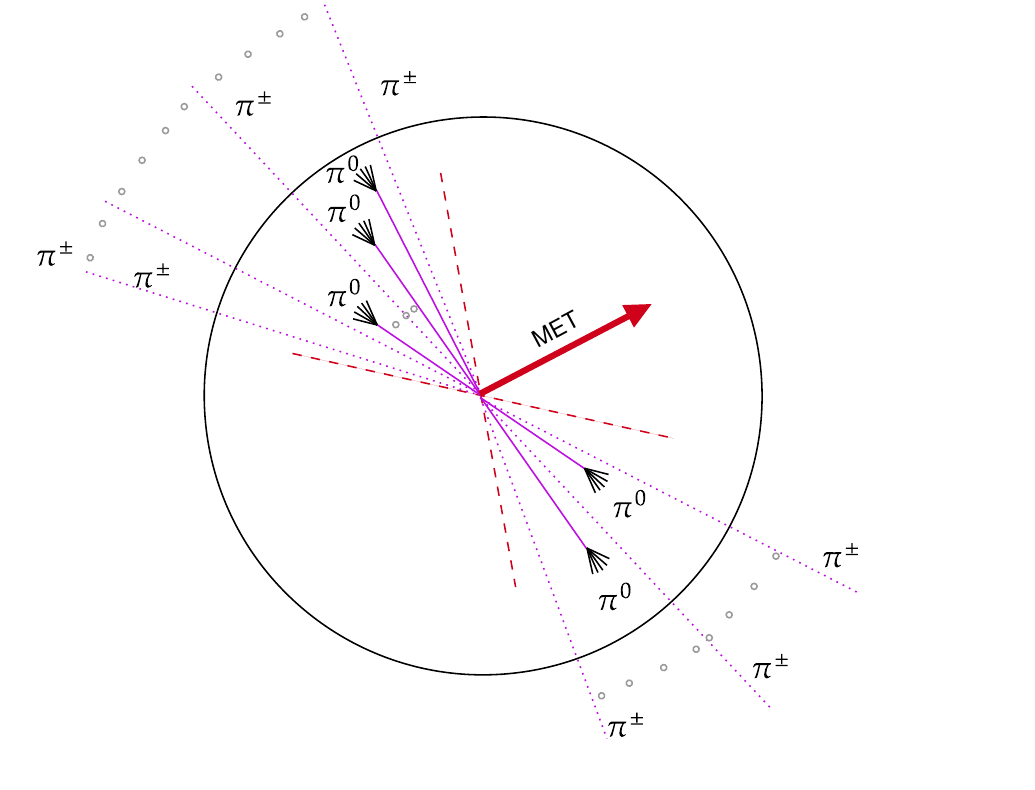}
    \caption{$2\to 2$ (left panel) and $2\to \rm{many}$ (right panel) production mechanisms in the detector plane. The  $2\to \rm{many}$ topology is of an interest to our SVEJ scenarios. We show an approximate jet cone (red dashed lines) within which the dark pions are produced, and we seperate them into  decaying diagonal pions (solid magenta lines) and stable off-diagonal pions (dotted magenta lines) which escape the detector undetected. The off-diagonal pions are stable by virtue of flavour symmetry.}
    \label{fig:schema}
\end{figure}

We establish characteristic features of such semi-visible emerging jet (SVEJ) scenarios and propose a new search strategy targeting dark shower production featuring multiple displaced vertices, as illustrated in Fig.~\ref{fig:schema}. It shows the approximate jet cone (red dashed lines) within which the dark pion production takes place. We differentiate between the diagonal pions (solid magenta lines) decaying to visible SM $u,d$ quarks and the stable off-diagonal pions (dotted magenta lines) which escape the detector. This illustrates our signal characteristics where a mixture of stable and long-lived pions gives rise to the signature of our interest.

Along with the analysis of this principle signature, we also discuss the reach of current long-lived particle searches for the same scenario to compare and contrast between different search strategies. With this work, we aim to demonstrate a possible strategy to classify experimental signatures and associated theoretical scenarios. 

The remainder of this paper is organised as follows. In sec.~\ref{sec:sig_details} we describe the details of our signal model, followed by generator level characteristics dictated by top-down approach in sec.~\ref{sec:gen_level}. This informs our discussion of analysis strategy described in sec.~\ref{sec:ESVJ_analysis} and we present corresponding results in sec.~\ref{sec:results} before concluding in sec.~\ref{sec:conclusions}. 
\section{Signal characteristics}
\label{sec:sig_details}
\subsection{Theory model}
As described previously, we consider an $SU(\nc)$ extension of the SM, where the mass-degenerate dark quarks charged under $SU(\nc)$ gauge group, are uncharged under the SM. In the chirally-broken phase, this theory contains the pseudo-Nambu Goldstone Bosons i.e. dark pions as the lightest degree of freedom. Alongside, this theory also contains heavier states such as $\rho, \sigma, a$ etc. In isolation, this HV/DS theory space consists of four free parameters. They are, number of flavors($\nf$), number of colors ($\nc$), one mass ratio and an overall mass scale. The mass ratio is typically chosen to be $\mpl$ -- the ratio of pion mass to the confinement scale of the theory, while the overall scale can be chosen as $\ld$. To be in the chirally broken phase the theory must contain $\fc \ll 3$, this is also the region where \textsc{Pythia} can be used~\cite{Kulkarni:2025rsl}. Additionally, we fix $\fc = 1$, as there is evidence that the relative dark pion and dark rho meson spectrum is $\nf$ dependent~\cite{Alfano:2025non}. From this point onward, we do not use subscript `D' to indicate dark pions or dark rho mesons. We will also refer to dark pions or dark rho mesons simply as pions or rho mesons and $\pi, \rho$ represent these dark states rather than the SM mesons.

The theory model we use is very similar to the one defined in~\cite{Liu:2025bbc}, which justifies our choice of using the pion lifetime as a free parameter. In contrast to that model, in this work we consider a $Z^\prime$ coupling only to the first generation quarks. The resulting Lagrangian can be written as 
\begin{eqnarray}
    \mathcal{L} &\supset& \mathcal{Q}_V^{\rm SM} \kappa_D   {Z^{\prime}}^\mu \bar{q} \gamma_\mu q + \mathcal{Q}_A^{\rm SM} \kappa_D   {Z^{\prime}}^\mu \bar{q} \gamma_\mu\gamma_5 q \nonumber  \\ 
    &+& \mathcal{Q}_V^D  \kappa_{\rm D}  {Z^{\prime}}^\mu \bar{q}_D \gamma_\mu q_D +\mathcal{Q}_A^D  \kappa_{\rm D}  {Z^{\prime}}^\mu \bar{q}_D \gamma_\mu\gamma_5 q_D \nonumber \\ 
    &+& Y_{ij} \phi_D \bar{q}_D q_D + (D_\mu \phi)^\dagger (D^\mu \phi),
\label{eq:lagrangian}
\end{eqnarray}
with $q = (u,d)$ and where $\mathcal{Q}_{V,A}^D$ and $\mathcal{Q}_{V,A}^{\rm SM}$ are $\nf\times \nf$ and $2\times 2$ the dark and SM $U(1)_D$ vector and axial-vector charge matrices respectively, $Y_{ij}$ is the dark quark Yukawa matrix associated with dark Higgs ($\phi$). Both the charge and Yukawa matrices are assumed to be real and diagonal for simplicity. $D_\mu$ is the appropriate $U(1)_D$ covariant derivative. $\kappa_D$ is the $U(1)_D$ gauge coupling. It should be noted that despite being diagonal $\mathcal{Q}_{V,A}^D$ can be chosen such that every diagonal entry is different, effectively destabilizing the diagonal pions.~\footnote{The expressions for pion lifetime are available in~\cite{Liu:2025bbc} and we do not repeat them here. It is worth stressing that the freedom in choosing the $\kappa_D$ coupling and the charge matrices allow us to treat the pion lifetime as a free parameter. } 

Finally, the existence of $U(1)_D$ axial-vector charges ensures that the HV/DS  $Z^{\prime}$ couples to both the SM and the dark fermions. Our choice of restricting to up and down quark couplings is motivated by the signature we consider. Coupling to second and third generation quarks lead to secondary displaced vertices in the signal which we do not target in our current analysis but plan to return to them in the future. Since we choose diagonal Yukawa and $U(1)_D$ charge matrices, the resulting diagonal HV/DS pions are unstable however the off-diagonal pions are flavour symmetry protected. For concreteness throughout this work, we let all diagonal pions decay, however we note that any number of diagonal pions may be made unstable depending on the charge assignments. The effectiveness of analysis strategy proposed in this work for such arbitrary number of decaying pions will be established in the future works. 

\subsection{Simulation setup}
Throughout this work we use long-term-stable version \textsc{MadGraph\_aMC@NLO v2.9.24} to generate hard process $pp \to Z^\prime \to q_D \bar{q}_D $ up to two extra SM partons. We pass these events through \textsc{Pythia 8.312} hidden valley module~\cite{Carloni:2010tw, Carloni:2011kk} to simulate HV parton shower, fragmentation and hadronization of the dark quarks~\footnote{We note here that a HV module is also available within the \textsc{Herwig} event generator, we use \textsc{Pythia 8} as it is publicly available.} and use MLM merging procedure. The hadronization parameters are set according to~\cite{Liu:2025bbc} and the {\tt ProbVec} parameter is fit according to the fit 3 also provided in~\cite{Liu:2025bbc}. We choose our $\mpl$ parameter such that $m_\rho > 2m_\pi$ and hence $\rho \to \pi \pi$ is allowed, therefore in our setup $\mpl < 1.4$ is obeyed. Therefore, in our setup both $\rho, \pi$ are produced and the $\pi$ multiplicity is affected due to $\rho \to \pi \pi$ decays. The details of this effect are discussed in~\cite{Liu:2025bbc}. We note here that it is also possible to obtain similar signatures in $\mpl > 1.4$ regime, via displaced decays of the dark $\rho$ mesons~\cite{Bernreuther:2025xqk}.

In the following discussions, unless otherwise specified, the events are generated using this setup and no detector simulation is employed. Our most basic objects are charged hadrons with $p_T > 1$ GeV within $|\eta| < 2.5$, which we refer to as tracks~\footnote{Since tracks with $p_T \gtrsim$ 400 MeV can be efficiently reconstructed~\cite{ATLAS:2017kyn}, we could have considered a looser selection. Here we are conservatively following the selection done in~\cite{ATLAS:2025bsz}, which aims at reducing the impact of pile-up.}. The track impact parameters are smeared both in the transverse ($d_0$) and longitudinal directions ($d_z$) using a Gaussian distribution. Following Section 4.4 from the CMS performance report~\cite{CMS:2014pgm}, we consider a constant resolution in the longitudinal direction $\sigma_z = 0.1$ mm while in the transverse direction we employ
\begin{equation}
\sigma_0 = \sqrt{(0.03)^2 + \left(\frac{0.01~\rm{GeV}}{p_T}\right)^2} ~\rm{mm} \, ,
\end{equation}
resulting for our softer tracks ($p_T= 1$ GeV) in $\sigma_0 \sim 0.032$ mm, while for hard tracks ($p_T \sim 100$ GeV) the $p_T$ dependent term has less relevance, resulting in $\sigma_0 \sim 0.03$ mm. 

We use and compare several different object definitions, specially those pertaining to jets, MET and $H_T$ as we consider different triggers and set our object definitions consistent with those. Object definitions are therefore specified at appropriate places throughout the discussions in following sections.
\section{Generator level event characteristics}
\label{sec:gen_level}

\begin{figure}
    \centering
    \includegraphics[width=0.49\linewidth]{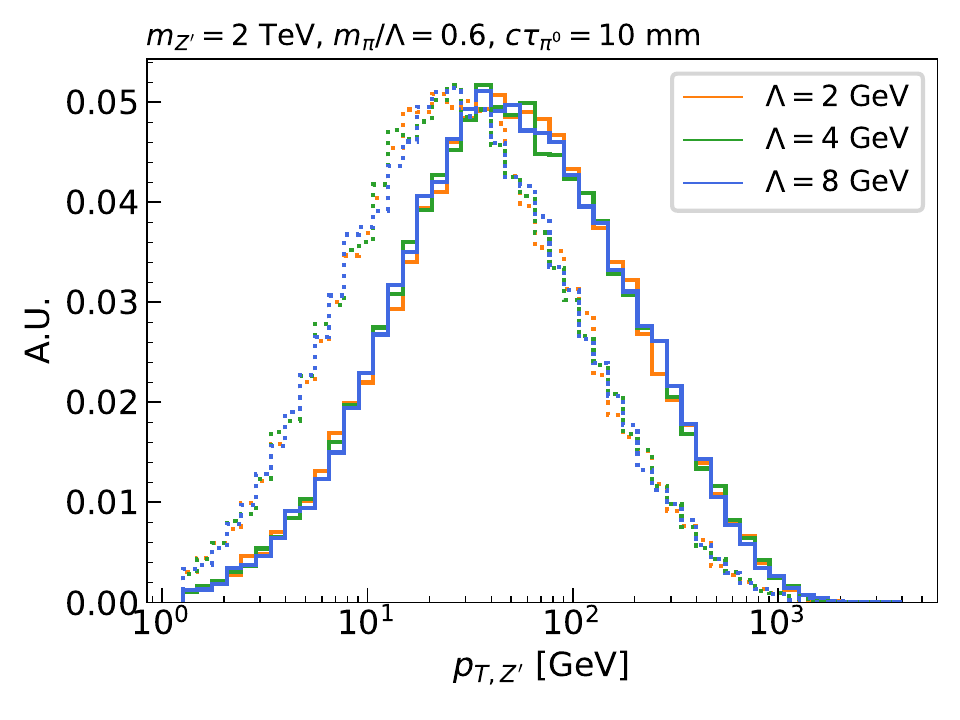}
    \includegraphics[width=0.49\linewidth]{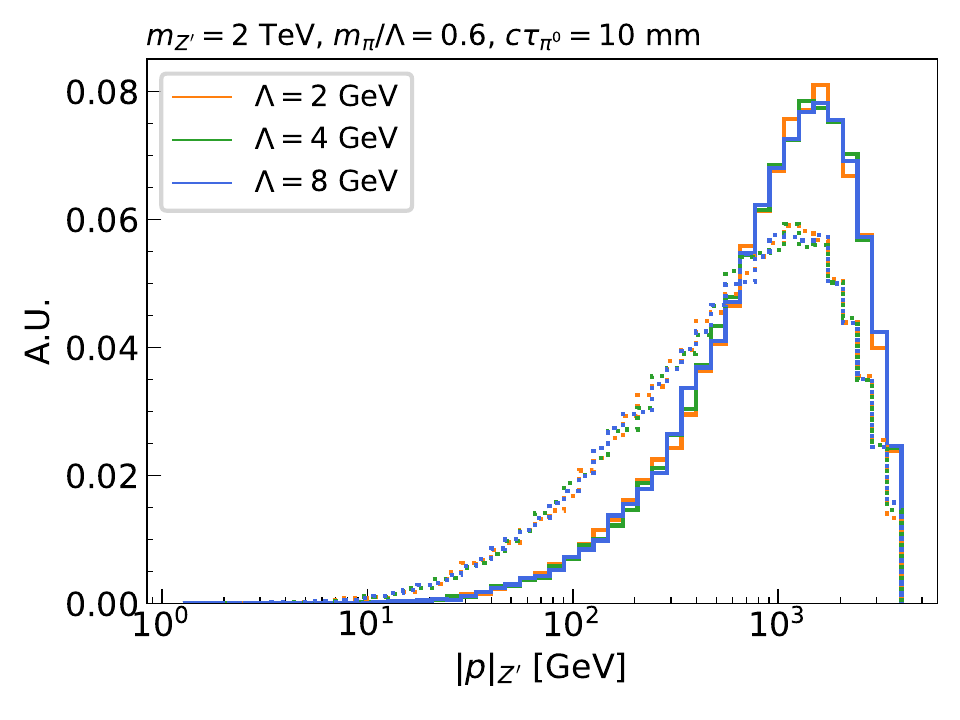}
    \caption{Distribution of the transverse and total momentum of the $Z'$ vector boson, with \textsc{Madgraph} generated events (solid lines), and \textsc{Pythia8} only generated events (dotted lines).}
    \label{fig:pT_zp}
\end{figure}

We begin by analyzing the characteristic kinematic distributions for the signal of our interest. Our discussion here motivates our choice of discriminating variables and serves as a crucial link between theory setup and corresponding experimental analysis design. We fix our benchmark as $\nc = 5, \fc = 1$, $\ld = 8, 50, 300$ GeV, with $m_\pi/\ld=0.6$, for concreteness\footnote{In the large-$\nc$ limit, the anomaly contribution to the singlet $\eta^\prime$ diminishes, which justifies our setup of treating $\pi$ and singlet $\eta^\prime$ to be mass-degenerate.}. Following the Snowmass fits~\cite{Albouy:2022cin}, this leads to $m_\rho=20.07, 250.1, 753$ GeV, and $m_{q_D}=0.09, 1.19, 3.57$ GeV, which are functions of $m_\pi/\ld$. Unless otherwise mentioned, we will consider $\mzp = 2\,\rm{TeV}$ and $\ctp = 10\,\rm{mm}$. As we will show later, the proposed SVEJ analysis has maximal sensitivity around $\ctp = 10\,\rm{mm}$, which justifies the choice of our benchmark lifetime.

In Fig.~\ref{fig:pT_zp}, we compare the total and transverse momentum distributions of the $Z^\prime$ for three representative values of $\ld$, obtained from matched and merged samples and from samples generated with the \textsc{Pythia8} Hidden Valley module alone. We see two important effects, first compared to \textsc{Pythia8} only production, the $Z^\prime$ receives additional boost in matched, merged samples. This is not surprising and it is a direct consequence of proper modeling of the hard initial-state-radiation. Second, this effect is independent of $\ld$ as it should be, since $\ld$ plays no important role in hard process as long as the partonic center-of-mass energy is much larger than $\ld$. Fig.~\ref{fig:pT_zp} thus justifies our choice of using matched-merged samples throughout this work. 

\begin{figure}
    \centering
    \includegraphics[width=0.49\linewidth]{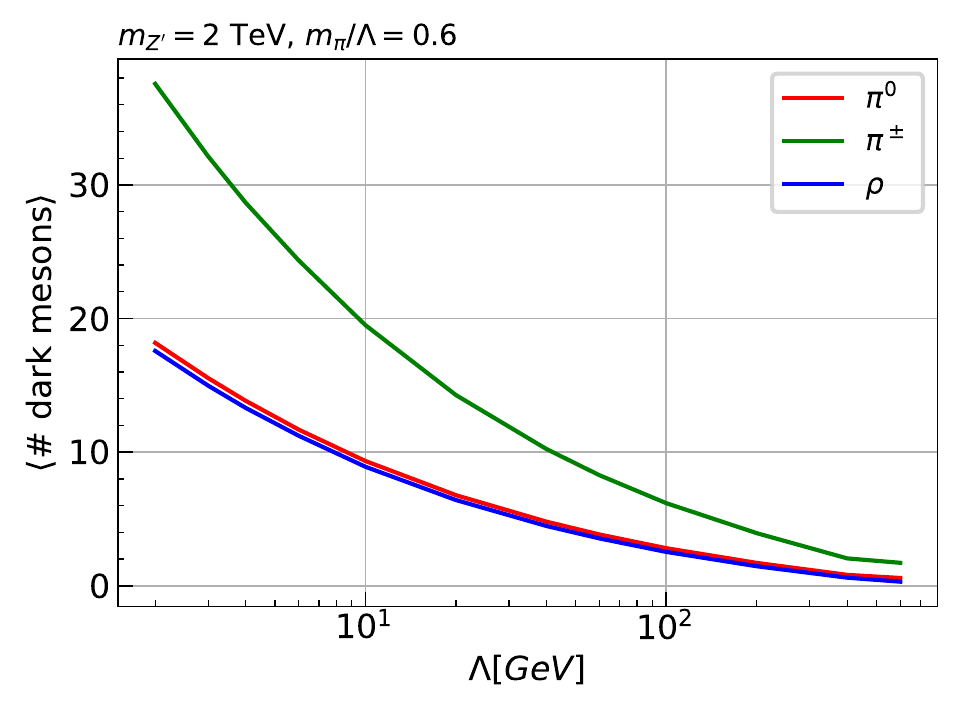}
    \caption{Averaged number of dark mesons per event as a function of $\ld$, for $\mzp = 2$ TeV and $\mpl = 0.6$ for several subspecies.}
    \label{fig:avg_ndark}
\end{figure}

In fig.~\ref{fig:avg_ndark}, we 
show the number of diagonal, off-diagonal pions and the total rho mesons as a function of $\ld$. These show a smooth distribution, inversely proportional to $\ld$ and shows a transition between $2\to$ many to $2\to$ few final states around $\ld = 100$ GeV. Therefore, for the final results section we restrict $\ld \leq 100$ GeV. Our choice $\ld \leq 100$ GeV is motivated by our interest in $2\to$ many final states. $2\to$ few final states may also be interesting however the validity of {\tt PYTHIA8} HV module to correctly simulate this regime should be established. 

\begin{figure}[h!]
    \centering
    \includegraphics[width=0.49\linewidth]{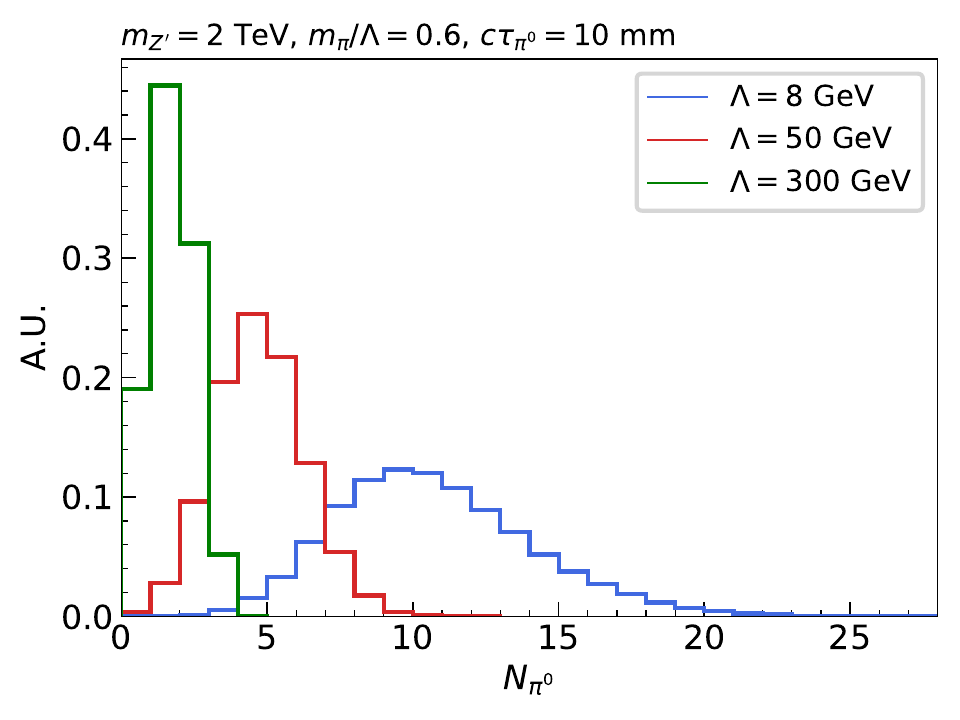}
    \includegraphics[width=0.49\linewidth]{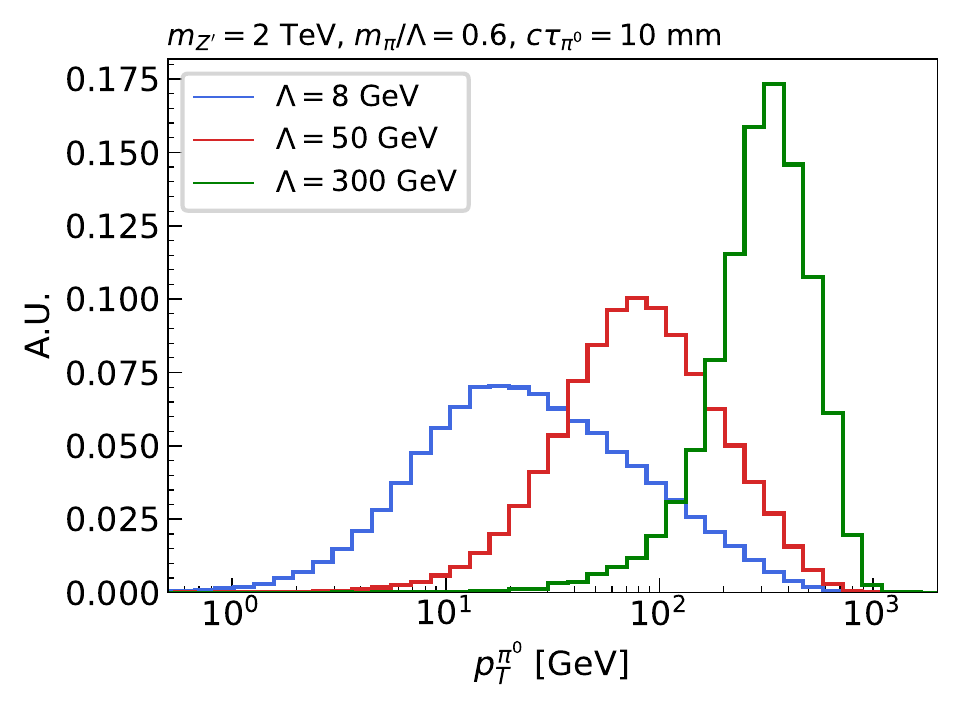}
    \includegraphics[width=0.49\linewidth]{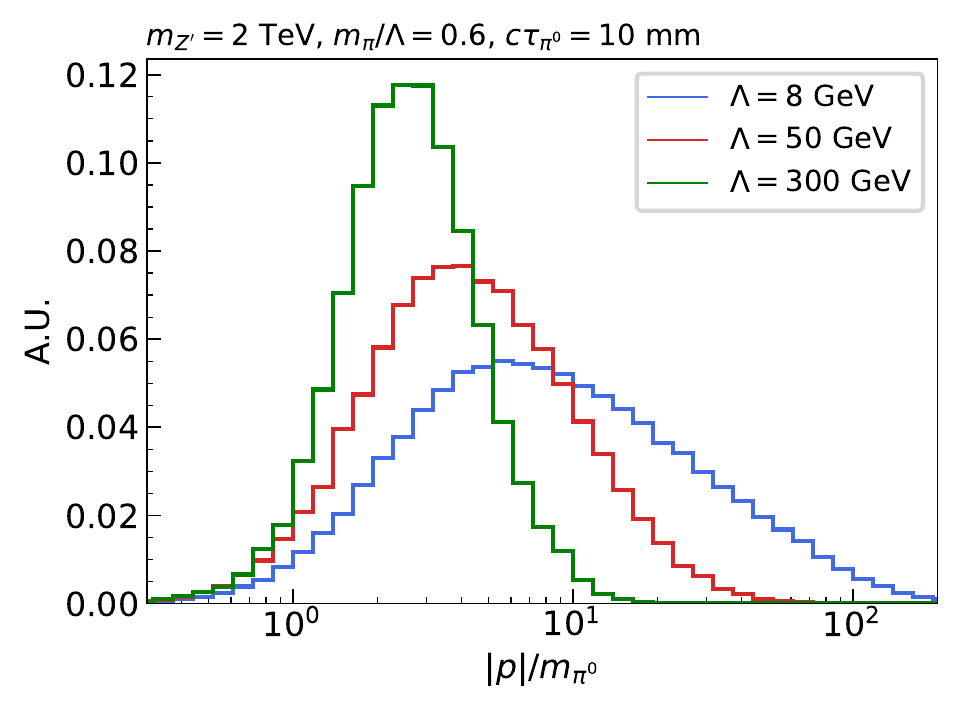}
    \includegraphics[width=0.49\linewidth]{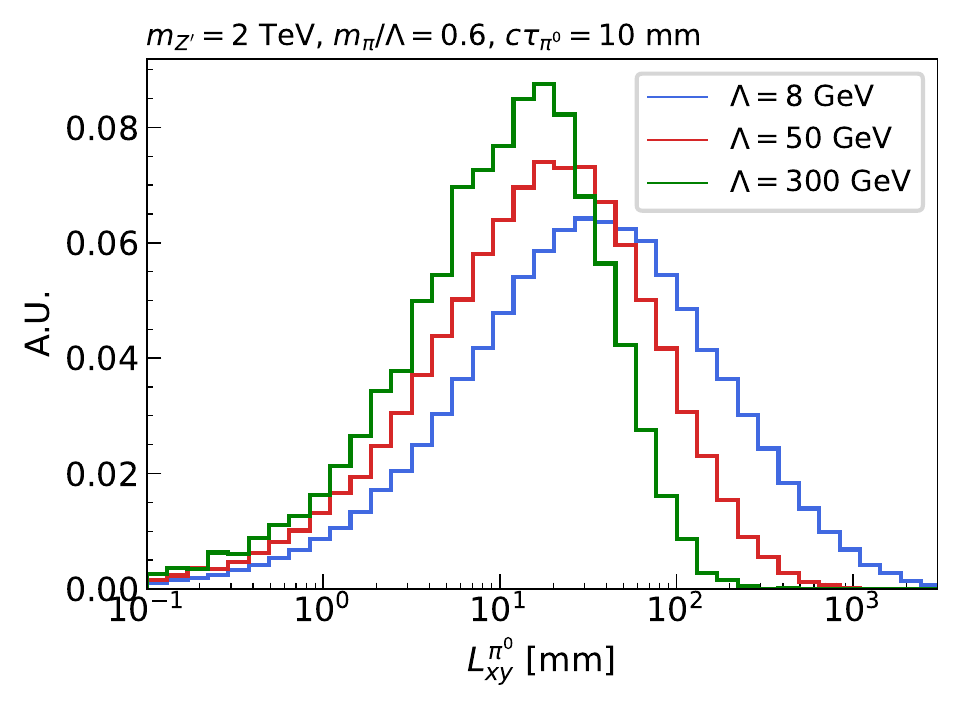}
    \caption{Normalized distributions of the number of the decaying dark pions $N_{\pi^0}$, their $p_T$, their boost factor and transverse decay length $L_{xy}$ for $\ld  = 8, 50, 300$ GeV, $\mpl = 0.6$, $\mzp = 2$ TeV and $\ctp = 10$ mm. }
    \label{fig:dist_kin_z2TeV}
\end{figure}

In fig.~\ref{fig:dist_kin_z2TeV}, we show several important kinematic distributions for the benchmark values of our signal model. We start by analyzing the number of diagonal i.e. unstable pions as a function of $\ld$. As expected the number of diagonal pions is largest for $\ld = 8\,\rm{GeV}$ given that the shower length is the longest. The number of diagonal pions decreases as $\ld$ increases and importantly for the highest value of $\ld = 300 \,\rm{GeV}$ the $N_{\pid}$ distribution peaks at 1. We also show that the $p_T$ of these diagonal pions is inversely proportional to $\ld$. 

\begin{figure}[h!]
    \centering
    \includegraphics[width=0.49\linewidth]{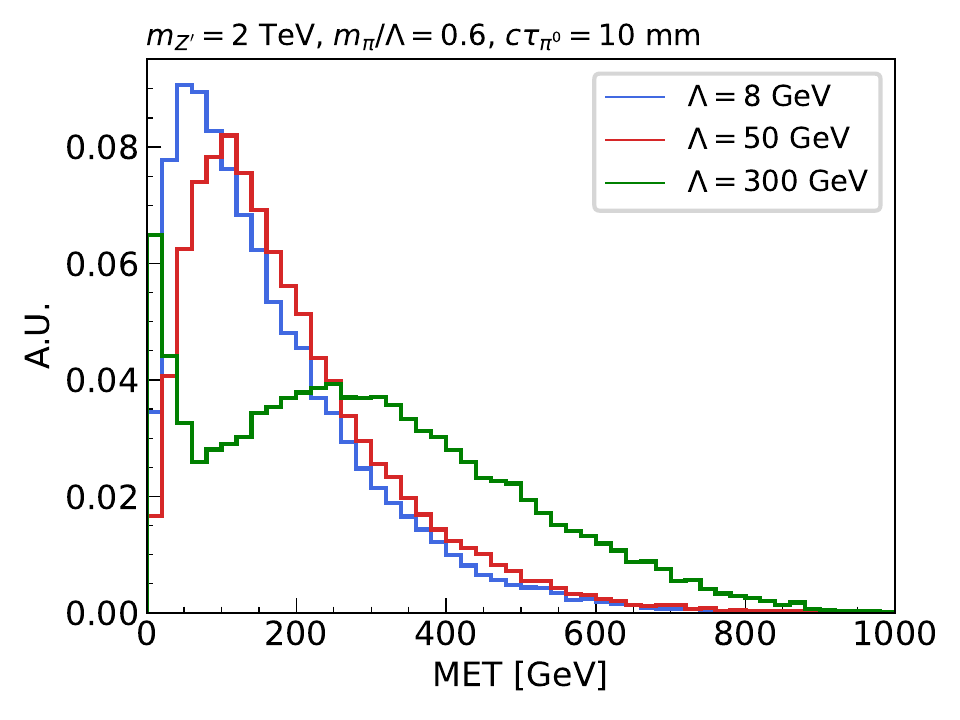}
    \includegraphics[width=0.49\linewidth]{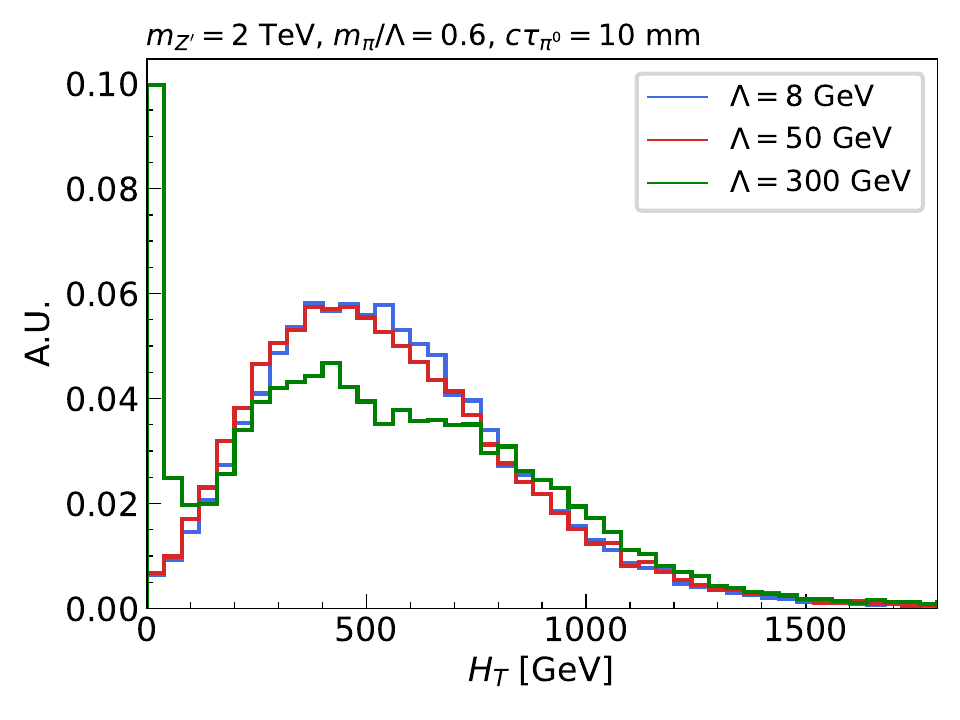}
    \caption{Missing energy (left panel) and the $H_T$ (right panel) distribution for a $\ld = 8, 50, 300\,\rm{GeV}$ with fixed $\ctp = 10\,\rm{mm}$, $\mzp = 2$ TeV, $\mpl = 0.6$. }
    \label{fig:met_HT_mzp_2TeV}
\end{figure}

In the bottom panel, we show the pion boost $p_T/\mpid$ and lab frame transverse displacement $L_{xy}$ for fixed proper length of $\ctp = 10\, \rm{mm}$. Given that, for a fixed $\mpl$, the pion mass is directly proportional to $\ld$, the lightest pions are more boosted and hence feature long transverse displacement as opposed to heavier pions. This has direct implications for the observable SM final states. First, we expect softer displaced vertices for lower $\ld$, and second, the value of $\ctp$ at which a given LHC search has maximum sensitivity depends on $\ld$ due to the change in the boost distribution. We will explicitly demonstrate this in the results section. 

\begin{figure}
    \centering
    \includegraphics[width=0.8\linewidth]{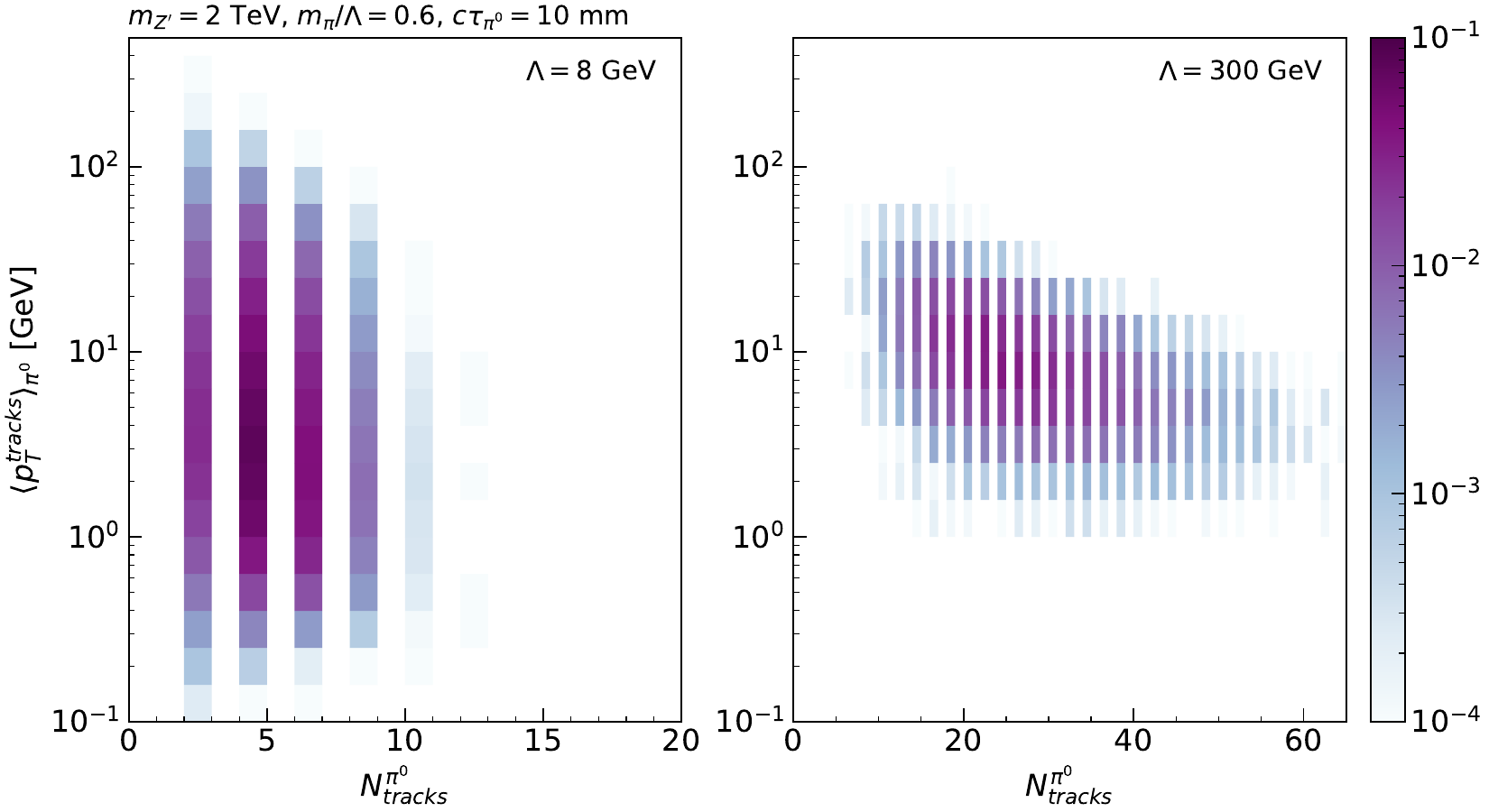}
    \includegraphics[width=0.8\linewidth]{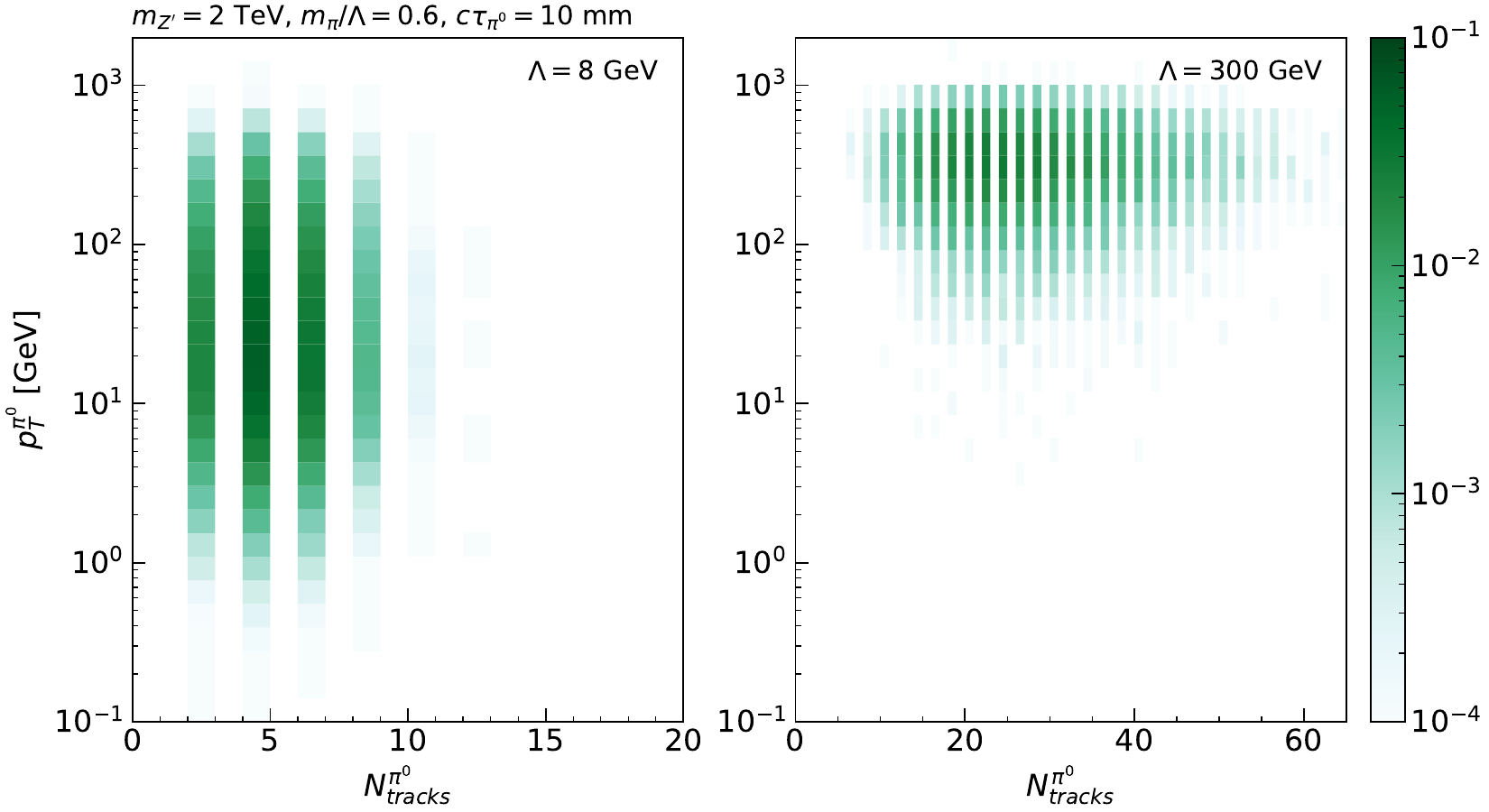}
    \caption{Normalized distribution of the number of tracks per pion (charged products) vs the average $p_T$ of the tracks (upper row) and the number of tracks per pion vs the pion transverse momentum (bottom row) for $\ld = 8\,\rm{GeV}$ (left panel) and $\ld = 300\,\rm{GeV}$ (right panel) for fixed $\mzp=2$ TeV, $\mpl = 0.6$ and $\ctp = 10\,\rm{mm}$. }
    \label{fig:trks-pion}
\end{figure}

In fig.~\ref{fig:met_HT_mzp_2TeV}, we show the missing energy (left panel) and the $H_T$ (right panel) distribution for $\ld = 8, 50, 300\,\rm{GeV}$ with fixed $\ctp = 10\,\rm{mm}$. These distributions will clear depend on $\ctp$, for example, if $\ctp \to \infty$, all pions will be stable at detector scale and the resulting signature will contain missing energy and initial-jet-radiation activity. The MET is computed as the modulus of the visible transverse momentum, MET = $|- p_T^{\rm vis}|$, by adding up the transverse momentum of electrons, muons and isolated photons with $p_T > 10 $ GeV and within $|\eta| < 2.5$, together with all final state hadrons with $p_T > 0.5$ and within $|\eta| <4.8 $. This implies that any particle decaying outside the detector is not contributing to the visible transverse momentum. The mean MET is comparable for $\ld = 8, 50$ GeV, however for $\ld =$ 300 GeV, the MET decreases strongly due to very few and heavier pions. This is also reflected in the $H_T$, which is defined as the scalar sum of all jets $p_T$, where jets are clustered using anti-$k_T$ algorithm with $\Delta R = 0.4$, and $p_T > 20$ GeV . The $H_T$ distribution in general features a long tail. Irrespective of the value of $\ld$ we find that it peaks at small $H_T$ values. The $H_T$ distribution peaks at zero for large $\ld$ because as shown in fig.~\ref{fig:avg_ndark}, the number of diagonal pions are very low which results in little visible activity particular if no initial-jet-radiation is present.

Finally, in fig.~\ref{fig:trks-pion}, we show the distributions for fraction of pions as a function of number of decaying pions and the average $p_T$ of tracks originating from those pions (top panel) as well as fraction of pions as a function of number of decaying pions and the average $p_T$ of pions (bottom panel) for fixed $\mpl = 0.6, \ctp = 10 \,\rm{mm}, \mzp = 2\,\rm{TeV}$ for two values of $\ld = $ 8 GeV (left panel) and 300 GeV (right panel). The difference in these distributions is striking. While on average the tracks per pion tend to feature small multiplicity and are soft for small $\ld$, they are harder and have a large multiplicity for larger $\ld$. This isn't surprising, but it underlines the importance of targeting soft but large track multiplicity environments. 

For completeness we comment on the CMS EJ search~\cite{CMS:2018bvr} applied to our s-channel mediator model, whose reinterpretation has been carried out in Ref.~\cite{Carrasco:2023loy}~\footnote{The associated code can be found in the LLP Recasting Repository~\cite{LLPrepo}.}. We find that the search efficiency is directly proportional to $\ld$. This implies that this search is not correctly targeting an s-channel emerging SVJ topology i.e. events with several dark pion production, and seems to favor a $2\to2$ rather than a $2\to$ many process (corresponding to an EJ topology). 

The above discussion therefore shows the following characteristics, first, the number of pions in an event decreases with $\ld$, while their $p_T$ is directly proportional. We note that for the 
 highest $\ld$ we consider, we often get no diagonal pions leading to MET dominated events, due to the expected multiplicity being below unity. This has implications for number of jets in an event, they are also a function of $\ld$ but curiously not a monotonically increasing or decreasing one. We expect the distribution of number of jets to non-trivially depend on $\ld$ and thus associated shower shapes. Looking at number of tracks therefore makes sense even at the cost of sensitivity to hadronization parameters. The number of tracks per pion show clear correlation with $\ld$, where for large $\ld$, both the number of tracks per pion and their $p_T$ is higher. Topologies featuring multiple dark pion production are therefore dominated by soft-displaced vertices. 

Taken together with the fact that our process inherently produces more than two pions in the final state, this motivates an analysis strategy based on binning by the number of decaying pions, that is, by the number of displaced vertices and tracks per vertex. We remind the reader here that either of these are highly sensitive to hadronization uncertainties, however as we describe below, our signal regions are defined in an inclusive manner to minimize this uncertainty. 
\section{Analysis strategy}
\label{sec:ESVJ_analysis}
Clearly, as our signal contains characteristics of both semi-visible and emerging jets, it is possible that a number of darkshower searches are applicable to the signal of our interest. Among them, ATLAS search~\cite{ATLAS:2025bsz} targets s-channel emerging jets. In fact, we employ the same trigger in our study but their event selection focuses on harder jets, with at least $p_T > 250$ GeV (see Table 2 in~\cite{ATLAS:2025bsz}), hence we expect it to be highly inefficient for our signal featuring multiple vertices with lower track $p_T$. Note that this search was designed having in mind a model where all (diagonal and off-diagonal) dark pions decay visibly, hence their model generates larger $p_T$ than in our scenario. Additionally, the semi-visible jet searches from both ATLAS and CMS~\cite{CMS:2021dzg, ATLAS:2023swa} exist, they focus on promptly decaying dark mesons. Finally, it is possible that mono-jet searches are applicable for dark mesons with very large lifetime but these are not the focus our investigation here. As an example of interplay of some of these searches for s-channel topologies see e.g.~\cite{Bernreuther:2019pfb}. We stress here that it will be highly interesting to examine this interplay for the model of interest here.

To design a search targeting our SVEJ topology (featuring several ``soft'' displaced vertices), we would need to conceive a strategy that includes a trigger from the current menu~\footnote{It is clear that the conspicuity of the final state could be leveraged with a dedicated trigger, but this study is beyond the scope of the current work.}, and an event selection that must not only be very efficient for signal, but also have a large background rejection. Regarding background, SM processes contribute here only a negligible fraction of the total background expected;  instead we must worry about background sources such as random track crossings, interactions with the material, tracks from cosmic rays, beam halo, etc, (for more details see e.g. Section 4 in~\cite{Alimena:2019zri}). A careful treatment of these backgrounds is clearly beyond the scope of this work, so here we will consider a set of simple cuts, inspired from a series of existing experimental studies. As we cannot reliably simulate these backgrounds, for the sensitivity estimations we will follow a simplified approach by considering scenarios with ``low" and ``moderate'' background counts, for $B=3$ and $B=100$ expected background events respectively. For 95 \% C.L exclusions, these curves are obtained by requiring the expected number of signal events $S$ to be 10 and 20, respectively. 

We will start by exploring first which of the existing triggers are suitable for our signal benchmarks in Section~\ref{s.triggers}, while in Section~\ref{s.signalregion} we will present a cut-based event selection aiming at reducing the backgrounds while keeping the signal as large as possible.
\subsection{Triggers}
\label{s.triggers}
We start by exploring, from the existing trigger menu, those that are employed in prompt searches (or ``standard triggers'' in the parlance of reference~\cite{CMS:2024zqs})~\footnote{For one or two physics prompt objects, Table 7 in~\cite{CMS:2024zqs} summarizes the corresponding thresholds.} and that will not obviously be highly inefficient for our signal of ``multiple soft displaced vertices''. Since our off-diagonal pions are stable, an obvious choice is to employ a missing transverse energy (MET) trigger, corresponding to MET $> 200$ GeV~\footnote{Note that in Table 7 of in~\cite{CMS:2024zqs} states the missing energy threshold to be 120 GeV; however this value corresponds to an online calculation; when considering the reconstructed (offline) missing momentum the selection is 95 \% efficient for $p_T{\rm miss} > $  250 GeV; we then quote 200 GeV as an optimistic,  improved threshold.}. However, from the left panel in figure~\ref{fig:met_HT_mzp_2TeV}, we do see that the MET follows a falling distribution. 

Along the same lines, we can also expect some hadronic activity from the decaying dark pions, as shown in the right panel of figure~\ref{fig:met_HT_mzp_2TeV}. Hence $H_T$
and its conjunction with MET, are also obvious candidates to try~\footnote{Here $H_T$ is ``the scalar pT
sum of all jets that meet certain selection criteria''~\cite{CMS:2024zqs}. We note that these criteria can vary from one analysis to another.}. As we will see in our summary, except for the inclusive MET trigger, their efficiency is at the few percent level, thus prompting the question of whether a more specialized trigger acting on displaced objects can have a large efficiency for our signal.

In a second step, we look into specific triggers targeting displaced jets (DJ). A recent CMS study~\cite{CMS:2024zqs} employs a new trigger based on $H_T > 430$ GeV together with two jets with $p_T > 40$ GeV within $|\eta| < 2$ and with at most one prompt track~\footnote{In this study, a prompt track is defined as a track with $p_T > 1$ GeV, transverse distance $d_{xy}$ to the primary vertex of at least 0.5 mm and a significance $d_{xy} / \sigma_{xy} < 5$, where $\sigma_{xy}$ is the uncertainty on $d_{xy}$.}. We will refer to this trigger as DJ, CMS. 

In a similar manner, ATLAS has deployed a series of triggers for Run-3 including two based on displaced jets and one specifically aimed to  emerging jets~\footnote{A summary of the Run-3 ATLAS trigger strategy can be found in~\cite{ATLAS:2024xna}.}. The ATLAS displaced jet triggers are based on counting the number of prompt and displaced tracks~\footnote{Tracks with $p_T > 1 $ GeV, within $|\eta| < 2.4$ and $\Delta R < 0.4$ of the jet as prompt 
(displaced) for $|d_0| < (>)~3$ mm.} within a given jet.  A prompt jet with $p_T > 180$ GeV is always required. This jet can be  accompanied by either (DJ-A, ATLAS) a single displaced jet with $p_T > 140$, $n_{pr} \leq 1$, $n_{disp} \geq 3$ or (DJ-B, ATLAS) a pair of displaced jets with $p_T > 50$ GeV, $n^{tr}_{prompt} \leq 2$, $n^{tr}_{disp} \geq 3$.

In addition, the recent ATLAS study on emerging jets~\cite{ATLAS:2025bsz} employs a dedicated trigger specifically designed for emerging jets (EJ, ATLAS). This trigger exploits the prompt track fraction (PTF) within a jet, a variable which is small for an emerging jet, and tends to unity for a QCD jet. Concretely, this trigger requires at least one large-R jet with $p_T > 200$ GeV, $|\eta| < $ 1.8 and PTF $<$ 0.08, where a prompt track must  be within $\Delta R = 1.2$ of the large-$R$ jet and satisfy $p_T > 1$ GeV, $|d_0|/\sigma_0 < 2.5$, $\delta_z = z_{\rm PV} - z_0 < 10$ mm, where $z_{\rm PV}$ is the position of the primary vertex in the longitudinal direction, and $z_0$ is the closest distance in the longitudinal plane between the track and the primary vertex.

\begin{table}[]
    \centering
    \small
    \begin{tabular}{|c|c|c|c|c|}
        \hline
        \multicolumn{5}{|c|}{Trigger efficiencies for $\mzp = 2$~TeV, $c \tau_{\pid} = 10$~mm} \tabularnewline
        \hline 
        Trigger & Ref &  $\ld =8$~GeV  & $\ld =50$~GeV  \\ 
        \hline
        MET($>200$ GeV) & ~\cite{CMS:2024zqs}  & 0.3150  & 0.3833 \\
        $H_T$ ($>1050$ GeV) &~\cite{CMS:2024zqs} & 0.0556  & 0.0568 \\
        MET-$H_T$ ($> 100 - 500$ GeV) &~\cite{CMS:2024zqs} & 0.3722 &  0.3765 \\
        \hline
        {\small DJ, CMS ($H_T>450$~GeV, $n_{DJ} \geq 2$, $n_{pr}\leq 1)$} &~\cite{CMS:2024xzb} & 0.3772  & 0.4288 \\
        {\small DJ-A, ATLAS ($p_T>180~{\rm GeV}, p_T^{DJ}>140 ~{\rm GeV}, n_{disp}\geq 3,n_{pr}\leq1)$} &~\cite{ATLAS:2024xna}  & 0.0671  & 0.1138 \\
        {\small DJ-B, ATLAS ($p_T>180$~GeV, $p_T^{1,2} > 50~{\rm GeV}, n_{disp}\geq 3,n_{pr}\leq 2)$} & ~\cite{ATLAS:2024xna}  & 0.0331  & 0.0846 \\
        \hline
        {\small EJ ATLAS ($p_T^{R=1} >200~{\rm GeV}, |\eta|<1.8, {~\rm PTF} < 0.08$)} & ~\cite{ATLAS:2024xna} & 0.6084  & 0.5500 \\
        \hline
    \end{tabular}
    \caption{Efficiencies for several triggers on our signal benchmark ($\mzp = 2$~TeV, $\ctp = 10$~mm), with  $\ld = 8, 50$~GeV. The second column points out to the reference where we obtained the trigger selection from.}
    \label{tab:triggers}
\end{table}

\begin{table}[]
    \centering
    \small
    \begin{tabular}{|c|c|c|c|c|}
        \hline
        \multicolumn{5}{|c|}{Trigger efficiencies for $\mzp = 1$~TeV, $c \tau_{\pid} = 10$~mm} \tabularnewline
        \hline 
        Trigger & Ref &  $\ld =8$~GeV  & $\ld =50$~GeV  \\ 
        \hline
        MET($>200$ GeV) & ~\cite{CMS:2024zqs}  & 0.0837  & 0.1227 \\
        $H_T$ ($>1050$ GeV) &~\cite{CMS:2024zqs} & 0.0068  & 0.0068 \\
        MET-$H_T$ ($> 100 - 500$ GeV) &~\cite{CMS:2024zqs} & 0.0615 &  0.0648 \\
        \hline
        {\small DJ, CMS ($H_T>450$~GeV, $n_{DJ} \geq 2$, $n_{pr}\leq 1)$} &~\cite{CMS:2024xzb} & 0.0656  & 0.080 \\
        {\small DJ-A, ATLAS ($p_T>180~{\rm GeV}, p_T^{DJ}>140 ~{\rm GeV}, n_{disp}\geq 3,n_{pr}\leq1)$} &~\cite{ATLAS:2024xna}  & 0.0302  & 0.0375 \\
        {\small DJ-B, ATLAS ($p_T>180$~GeV, $p_T^{1,2} > 50~{\rm GeV}, n_{disp}\geq 3,n_{pr}\leq 2)$} & ~\cite{ATLAS:2024xna}  & 0.0126  & 0.0180 \\
        \hline
        {\small EJ ATLAS ($p_T^{R=1} >200~{\rm GeV}, |\eta|<1.8, {~\rm PTF} < 0.08$)} & ~\cite{ATLAS:2024xna} & 0.2062  & 0.1848 \\
        \hline
    \end{tabular}
    \caption{Efficiencies for several triggers on our signal benchmark ($\mzp = 1$~TeV, $\ctp = 10$~mm), with  $\ld = 8, 50$~GeV. The second column points out to the reference where we obtained the trigger selection from.}
    \label{tab:triggersz1}
\end{table}

The efficiencies of all the triggers described above, for our signal benchmarks  with $\ctp = 10$ mm and $\mzp = 2$ TeV are shown in Table~\ref{tab:triggers}, while for $\mzp=1$ TeV those are displayed in Table~\ref{tab:triggersz1}. These efficiencies are computed as the ratio of the Monte Carlo events passing a specific requirement, normalized with respect to the number of Monte Carlo events after the matching/merging procedure~\footnote{We initially requested $10^5$ events at parton level in Madgraph, and after matching/merging on Pythia we keep on average 86 \% (73 \%) of those events for the $\mzp=1$ (2) TeV sample.}.

From these tables we see that from the ``standard'' triggers, MET is the most efficient (due to the lowest threshold), while the DJ triggers have a lower efficiency for $\mzp = 1$ TeV case and slightly higher for the $\mzp=2$ TeV case. The ATLAS EJ trigger is better suited for our purposes, with an efficiency of about 55-60 \% for our 2 TeV benchmark, and about 20 \% for the 1 TeV case.

\begin{figure}
    \centering
    \includegraphics[width=0.49\linewidth]{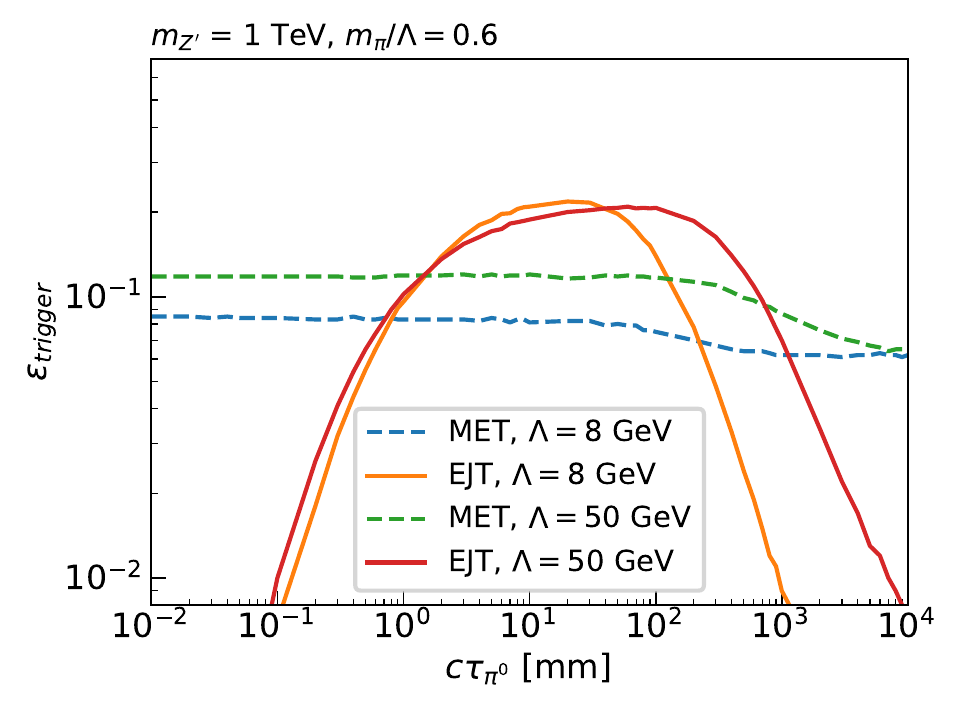}
    \includegraphics[width=0.49\linewidth]{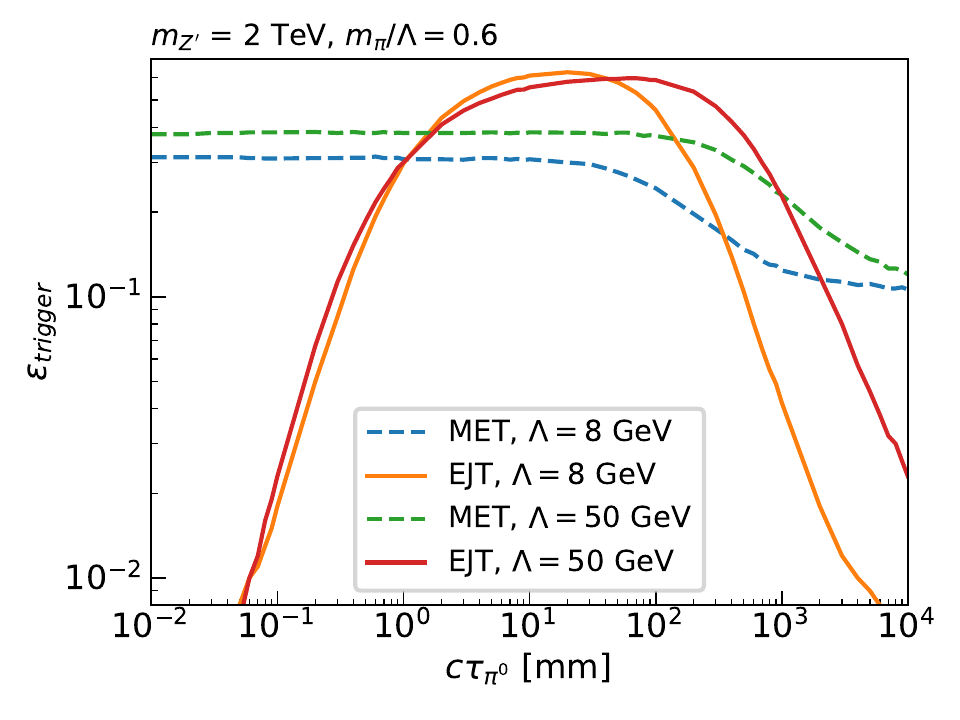}
    \caption{MET and EJ trigger efficiencies as a function of $\ctp$, for $\mzp =1$  TeV (left panel) and $\mzp =2$ TeV, for $\Lambda=8, 50$ GeV.} 
    \label{fig:eff_trig_ctau}
\end{figure}

Given that MET and EJ triggers stand-out for our benchmark point~\footnote{The DJ, CMS trigger is also promising, but since it depends on reconstructing $R=0.4$ jets its efficiency could vary depending on the hadronization parameters in the dark sector. In contrast, the EJ trigger from ATLAS uses fat-jets ($R=1$) is expected to be more robust against these hadronization parameters.}, it is interesting to consider the trigger efficiency dependence with the dark pion lifetime. We display them in figure~\ref{fig:eff_trig_ctau}, for our benchmark with $\mzp = 2$ TeV and $\Lambda=8, 50 $ GeV.

From the figure we see that, as expected, the EJ trigger is ideally suited to explore the lifetime range where the decays tend to occur in the tracker; the precise range depends on the dark pion boost, which is larger for lower $\Lambda$. but in both cases the EJ trigger is more efficient than MET for ${\cal O}(1~\rm{mm})$, and then for $\Lambda=8$ (50) GeV the efficiencies become similar for $\ctp = 100$ (1000) mm. The MET trigger behaves as expected. In our setup only the diagonal pions decay (giving rise to the visible decays that are used to compute MET), but there are more off-diagonal pions, which are fully stable. This imbalance in the number of decaying versus stable particles (all with the same mass scale) naturally generates a non-zero MET. For very low lifetimes, all the diagonal pions decay promptly, hence the MET shape is fixed solely by $\mzp$ and $\sqrt{s}$, and the MET distribution does not change. For large lifetimes, we see another plateau: now only few to no pions decay, and hence the resulting initial state radiation would contribute significantly to the MET; the intermediate range shows the transition between the two regimes.

\subsection{Event selection}
\label{s.signalregion}
We define displaced vertices as any vertex with three or more tracks associated to it, with our tracks fulfilling the conditions described in Section~\ref{sec:gen_level}. In addition, we only consider \emph{displaced} tracks, namely those fulfilling $|d_0| > 0.1$ mm, as the displaced vertex reconstruction employed by ATLAS in~\cite{ATLAS:2024qoo} (whose selection we largely followed) only considers such tracks.

We employ a simplified version of the vertexing algorithm, described in Ref.~\cite{ATLAS:2024qoo}. A crucial ingredient to greatly reduce the computation time is to employ the truth-level information on the track origin as our candidates for DVs. This procedure is justified because, when simulating only signal events, random crossings within a single event are unlikely to produce fake vertices due to their topology. By looping over the track collection sorted by decreasing transverse momenta, we attach all tracks to a given vertex if their origins are closer than 0.1 mm; for simplicity consider the DV position as the hardest track origin. To rule out material interactions, we only keep vertices whose positions do not coincide with dense regions of the detector. The thus-defined fiducial volume (FV) corresponds to longitudinal $z_{DV}$ and transverse position $\rho_{DV}$ of the displaced vertex below 300 mm, and $\rho_{DV}$ ought to be more than 1 mm away from any pixel layer~\footnote{Here for simplicity we follow the ATLAS layout, hence $\rho^{\rm layer} = 33.3, 50.5, 88.5, 122.5$ mm.}.

While these selections are, in principle, highly efficient for our signal, we supplement them with tighter requirements on the vertices, to reduce background. To remove vertices from pile-up, we require that at least one track has to fulfill $|d_0 / d_Z| > $ 0.25, (following Ref~\cite{CMS-PAS-EXO-24-033}), and each vertex must contain at least a track with $|d_0| > 3$ mm. To further reduce background, two additional cuts (taken from~\cite{ATLAS:2025bsz}) are imposed: i) the largest angular distance between any pair of tracks in the vertex, $\Delta R$ must fulfill $m_{DV} / \Delta_R >  4 {\rm~GeV}$, where $m_{DV}$ is the DV invariant mass and ii) the scalar sum $p_T$ of all tracks associated to a given vertex, $\Sigma p_T^{DV}$ has to be larger than 10 GeV. 

We note that each DV candidate will be defined by a number of tracks associated to it, hence when counting DV the minimum track requirement must be specified, $N_{DV}^{i} = N,i$ is then the number of events with $DV$ number of displaced vertices that have at least $i$ tracks associated to each DV. We note that in standard displaced vertices analysis a minimum of 5-tracks per DV are required (see e.g.~\cite{ATLAS:2017tny,CMS:2018tuo,CMS:2021tkn}). Needless to say that 2 tracks would be the bare minimum, but one would also expect a very large background from e.g. random crossings. In this respect, the analysis carried out in ~\cite{CMS:2021tkn}, where results are shown for vertices formed from 3 and 4 tracks~\footnote{It is worth noting that in this study the 3-track and 4-track vertices are not used for the signal region, but rather for the control region. DVs within the signal region are also requested to have five or more tracks.}, was an important guide for our procedure. From this analysis, we see that a large background reduction, for 3 (4) tracks attached to a vertex, is obtained if one requires a minimum transverse distance between the vertices, $d_{vv} > 1.5~(1)$ mm~\footnote{We stress that our study is targetting a different kinematic phase space; hence we make no claim of being background-free, but rather that such a cut would be helpful in further reducing the background.}.

The detailed cutflow from these cuts on our previously studied benchmarks is displayed in Table~\ref{tab:cutflow_MM}, for the $3,3$ signal region. The first row displays the EJ trigger efficiency, being the same as quoted in Table~\ref{tab:triggers}. 

\begin{table}[]
\renewcommand{\arraystretch}{1.2}
    \centering
    \begin{tabular}{|c|c|c|c|c|}
        \hline
        $\ctp = 10$ mm 
        & \multicolumn{2}{c|}{$\mzp=1$ TeV} & \multicolumn{2}{c|}{$\mzp=2$ TeV }
        \\
        
         Cutflow & $\Lambda = 8$ GeV & $\Lambda = 50$ GeV & $\Lambda = 8$ GeV & $\Lambda = 50$ GeV 
         \\
        \hline
	      EJ-trigger & 0.2062 & 0.1848 & 0.6084 & 0.5500 \\
        $N_{DV}\geq3$ ($\geq3$ trks) in FV & 0.1291 & 0.1172 & 0.4911 & 0.4733 \\
        1 trk, $|d_0/d_z|<0.25$ & 0.1288 & 0.1159 & 0.4906 & 0.4712 \\
        1 trk, $|d_0|>3$mm & 0.0821 & 0.0700 & 0.3619 & 0.3524 \\
        $m_{DV}/\Delta R>4$ GeV  & 0.0582 & 0.0658 & 0.2906 & 0.3387 \\
        $\Sigma p_T^{DV}>10$ GeV & 0.0527 & 0.0652 & 0.2712 & 0.3366 \\
        $d_{vv}>1/1.5$ mm & 0.0526 & 0.0646 & 0.2698 &  0.3322 \\
        \hline
    \end{tabular}
    \caption{Cut-flow efficiencies for signal of $N_{DV}\geq3$ (with $\geq3$ tracks) for different values of $\ld$ and $\mzp$.}
    \label{tab:cutflow_MM}
\end{table}

We see that the biggest impact on the signal occurs for two requirements: asking for a minimum of 3 vertices in the fiducial volume and demanding a highly displaced ($|d_0| > 3$ mm) track within a vertex. The requirements on $m_{DV} / \Delta_R$ and $\Sigma p_T^{DV}$ (which are informative on the boost of the dark pion), impact the signal events for low $\Lambda$, but are less important for the large $\Lambda$ case. The remaining cuts tend to have a minimal impact on the signal event. Regarding the variation with the lifetime, it can be anticipated from our discussion on $\Lambda$, hence we will not show the explicit results here. 

\begin{figure}
    \centering
    \includegraphics[width=0.49\linewidth]{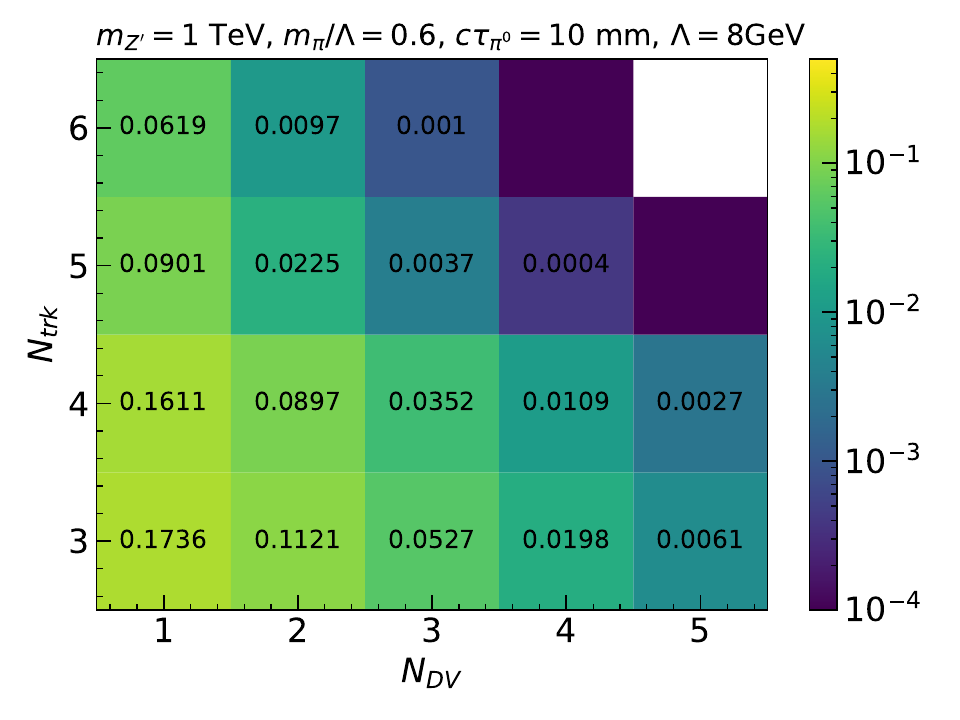}
    \includegraphics[width=0.49\linewidth]{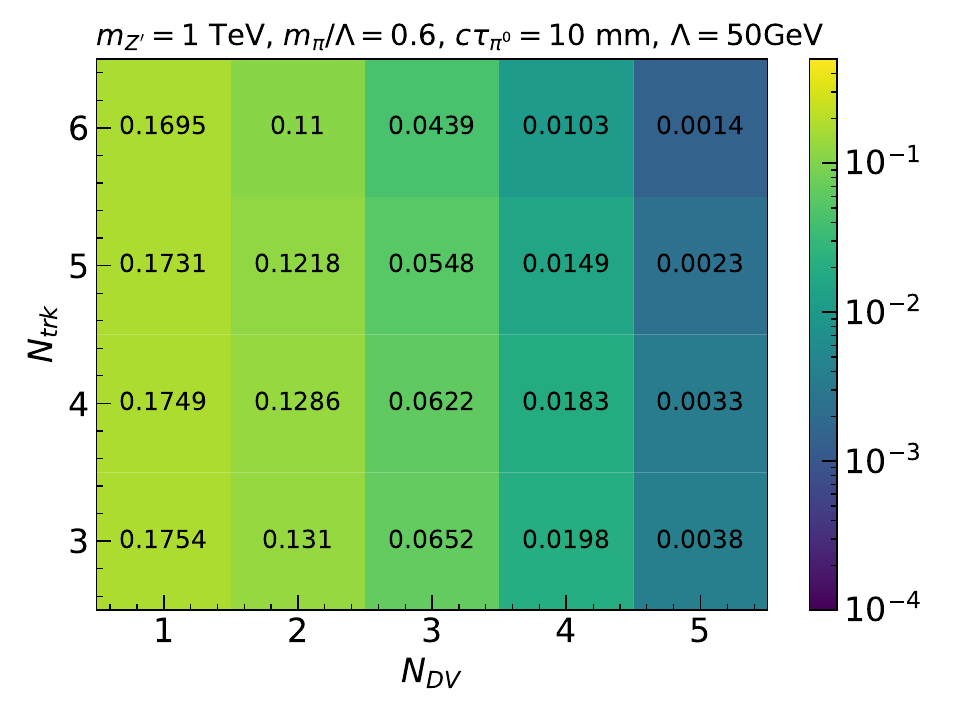}
     \includegraphics[width=0.49\linewidth]{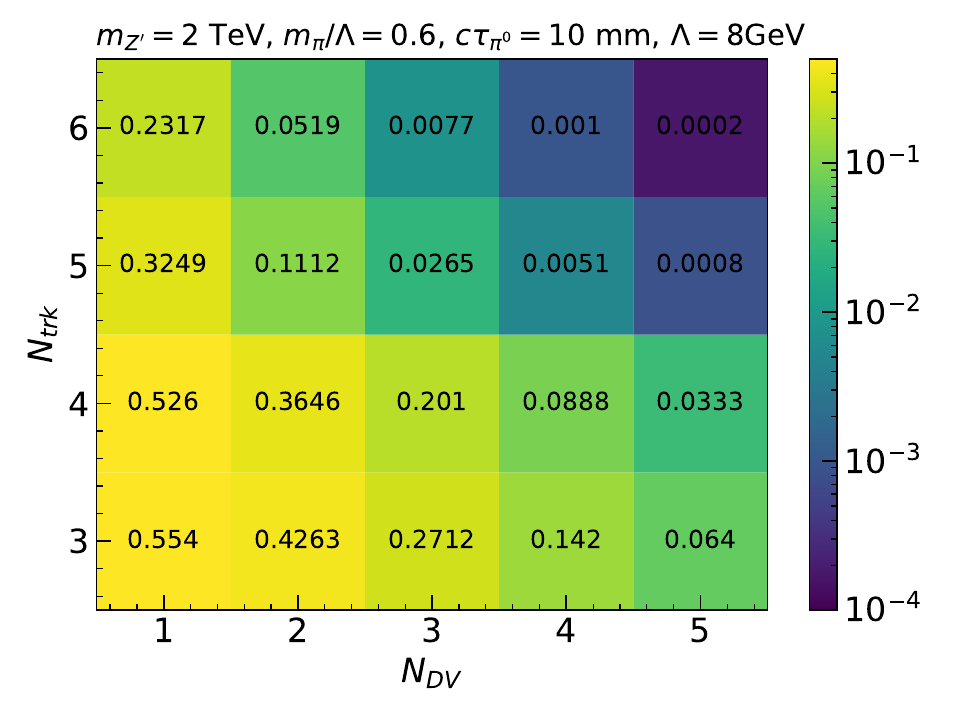}
      \includegraphics[width=0.49\linewidth]{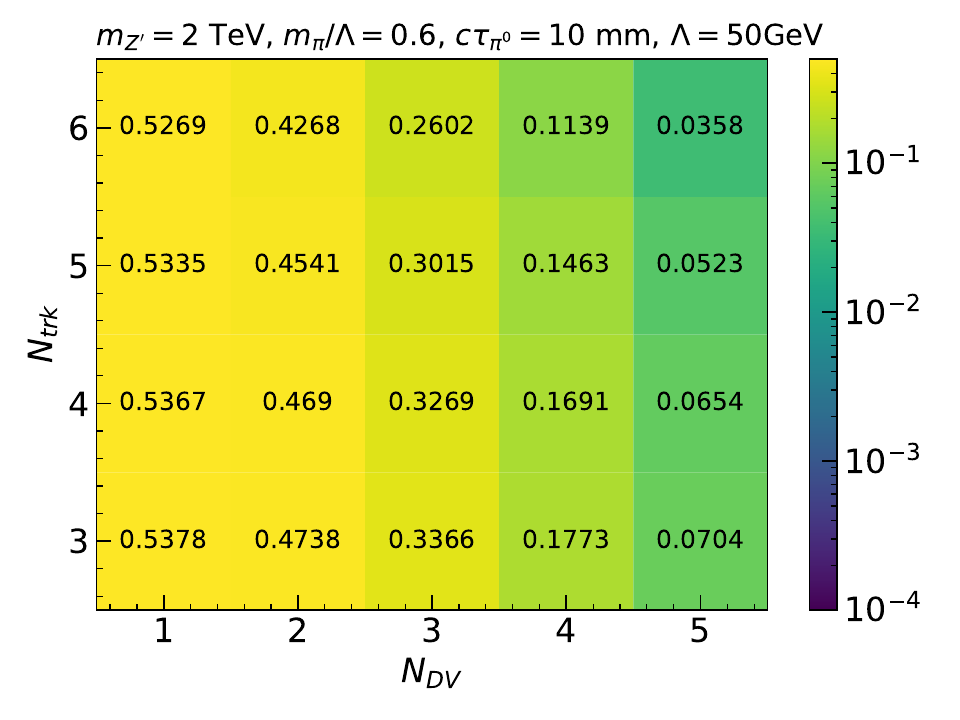}
    \caption{Distribution of the number of DVs per event against the number of tracks per DV, after the EJ-trigger and all DV cuts (except $d_{vv}$), for $\Lambda=8, 50$ GeV considering inclusive bins, for $\mzp=$1 (upper) and 2 TeV (lower) and $\ld = 8$ (left panel) and 50 GeV (right panel). }
    \label{fig:ndv_ntrk}
\end{figure}

It is also interesting to consider the variation on the number of tracks and number of vertices required in the study. We show in figure~\ref{fig:ndv_ntrk} the cutflow efficiency (where we have included all the previous cuts, except for the $d_{vv}$ requirement) when $N_{DV}$ vertices with at least $N_{trk}$ tracks are requested.

From the distribution we see that the $3,3$ region has an efficiency of about 34 \% for $\mzp=2$ TeV and $\Lambda=8$ GeV, which decreases to 28 \% when considering a larger $\Lambda$, and it lowers to $5-6 \%$ for the $\mzp = 1$ TeV case. These figures also quantifies the intuition that an extra track is ``cheaper'' than an extra vertex, as the $3,4$ region is larger than the $4,3$ one. 

It is worth now commenting on the usefulness of the $d_{vv}$ cut. In practice we see (last row from Table~\ref{tab:cutflow_MM}) that this latter cut is almost 100 \% efficient for our signal benchmark (and pretty much independently of the lifetime and the dark pion mass). Hence it is also clear that from our rationale to take more than 2 DVs each with more than 2 tracks in an inclusive manner, our most sensitive signal region would be $3,3$. Of course, it is a conceivable scenario that this $3,3$ region for $N>2$ could have some background at HL-LHC, while requiring a large number of tracks ($N_4^{(3)}$ or $N_3^{(4)}$) could give ``low'' background, hence changing the preferred signal region; properly assessing this situation requires a careful analysis by the experimental collaborations, probably exploiting a data-driven background extraction, and including realistic track reconstruction efficiencies.

We also note that, unlike the existing experimental studies, here we do not perform any association of the displaced vertices to jets; our proposed analysis is based entirely on the charged tracks forming vertices. Needless to say such an association would of course be important to pinpoint the signal features, but operating at a simpler level, and given the conspicuity (mostly in terms of multiplicity) of our signal, the vertex-based analysis is interesting enough to be sensitive to poorly constrained regions of parameter space.

\begin{figure}
    \centering
    \includegraphics[width=0.49\linewidth]{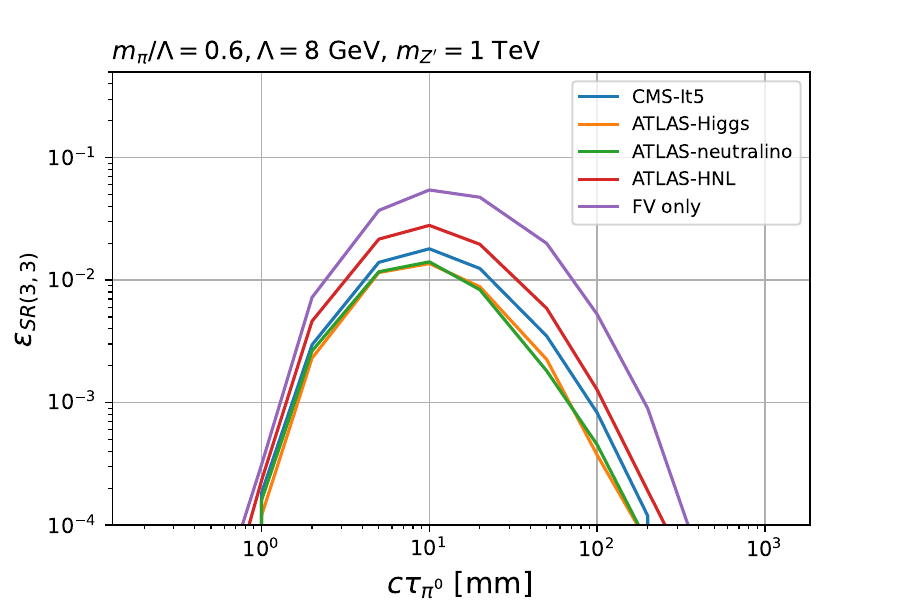}
    \includegraphics[width=0.49\linewidth]{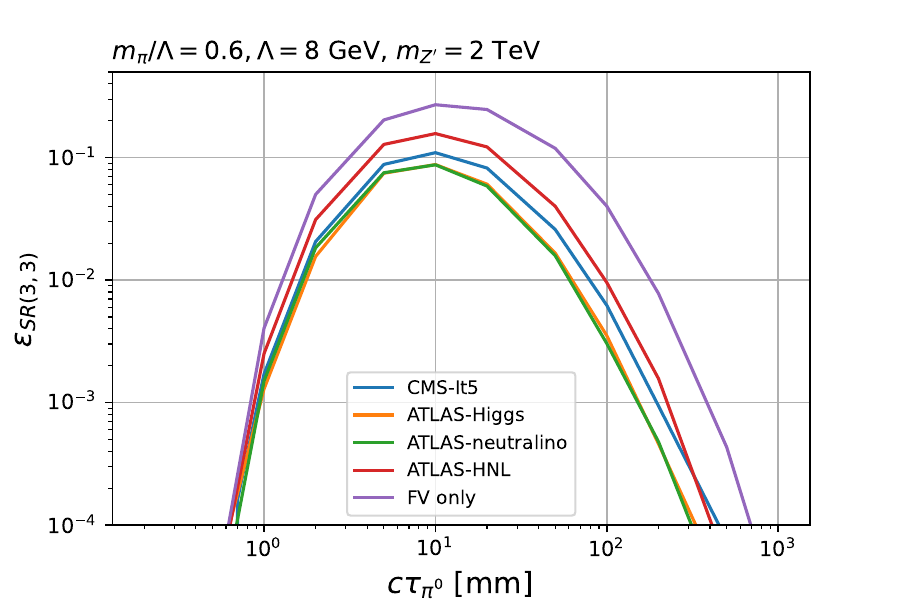}
    \caption{Cut-flow efficiencies for the $3,3$ signal region, as a function of $\ctp$, for $\mzp =1$  TeV (left panel) and $\mzp =2$ TeV, for $\Lambda=8$ GeV, under several assumptions for the tracking efficiency $\epsilon_{\rm trk}$. See main text for details.} 
    \label{fig:track_eff}
\end{figure}

Until now, we have been optimistic about the track reconstruction efficiency $\epsilon_{\rm trk}$ , keeping it at 100\%; this is a simplifying assumption and obviously not completely realistic. While the experimental collaborations have provided some simplified $\epsilon_{\rm trk}$ plots as function of a few kinematic variables (see e.g.~\cite{ATLAS:2023nze} for ATLAS and~\cite{CMS:2014pgm} for CMS), we acknowledge that such a simple parametrisation might not fully capture the complex dependence of this quantity on the underlying kinematics. This impossibility  supports our simplified choice of simply having 100~\% tracking efficiency in our study. Nonetheless, to have a rough idea of the impact of a realistic  $\epsilon_{\rm trk}$  we show the signal selection efficiency under several assumptions for $\epsilon_{\rm trk}$ in figure~\ref{fig:track_eff}, for two different values of $m_Z'$, as a function of the dark pion lifetime.

In the figure we compare the $\epsilon_{\rm trk} = 1$ result (purple) to that from CMS following the ``Iteration 5'' stage (figure 12 from~\cite{CMS:2014pgm} (blue), together with the efficiencies reported by ATLAS (Figure 7 of~\cite{ATLAS:2023nze}). The ATLAS collaboration have presented these tracking efficiencies only for three different BSM signal scenarios (Higgs Portal, neutralino, Heavy Neutral Leptons - HNLs) as a function of radial decay position. As none of these models matches our signal topology (the closest to our model is the Higgs Portal as it is a s-channel model, but there is a huge difference between having the center-of-mass energy at the SM Higgs mass instead of laying in the multi-TeV range), we decided to use the three of them to illustrate the systematic effects. All in all, we can conclude that for low displacements (low lifetime) all the scenarios tend to coincide, yet for large displacements there is a loss of efficiency due to the poorer reconstruction of tracks with larger production radius. At the position where the signal efficiency peaks this reduction is about a factor of 2-4 (from the ideal case), while for larger lifetimes one sees up to an order of magnitude in reduction for both the 1 and 2 TeV Z' mass benchmarks. Note that the efficiency differs between the different parametrisations by about a factor of two along the lifetime range of interest (with ATLAS-HNL being the case closest to ideal and the ATLAS-neutralino and ATLAS-Higgs portal being quite similar and the yielding the largest loss in signal efficiency. The fact that the CMS parametrization lies in between the different signal scenarios employed by ATLAS largely stresses the strong model-dependence of $\epsilon_{\rm trk}$. We conclude that while a proper assessment of the tracking efficiency is necessary for our model, including it in this study is beyond the scope of this work. 

It is also interesting to comment on the variation with $N_F$. We note first that the {\tt PYTHIA8} HV module has only been validated for $N_F=8$. By fixing $N_C=5$ and varying $N_F$ between 2 and 8, we have verified that the impact on the trigger efficiencies and on the cutflow for the (3,3) selection is minor, which renders our search robust against variations on $N_F$.

\begin{table}[]
\renewcommand{\arraystretch}{1.2}
    \centering
    \begin{tabular}{|c|c|c|c|}
        \hline
        $\ctp = 10$ mm & \multicolumn{3}{c|}{$\mzp=2$ TeV }\\
        
         Cutflow & $\Lambda = 8$ GeV & $\Lambda = 8/3$ GeV & $\Lambda = 24$ GeV 
         \\
        \hline
	      EJ-trigger & 0.6084 & 0.6160 & 0.6023\\
        $N_{DV}\geq3$ ($\geq3$ trks) in FV & 0.4911  & 0.5517 & 0.5796 \\
        1 trk, $|d_0/d_z|<0.25$ & 0.4906  & 0.4451 & 0.5331
        \\
        1 trk, $|d_0|>3$mm & 0.3619 & 0.4443 & 0.5326
        \\
        $m_{DV}/\Delta R>4$ GeV  & 0.2906  & 0.3078 & 0.4290
        \\
        $\Sigma p_T^{DV}>10$ GeV & 0.2712  & 0.2407 & 0.3548
        \\
        $d_{vv}>1/1.5$ mm & 0.2698 &  0.2230 & 0.3336
        \\
        \hline
    \end{tabular}
    \caption{Impact of hadronization uncertainties on the cutflow efficiency (for signal of $N_{DV}\geq3$, with $\geq3$ tracks) through $\ld$ variation while keeping the mass spectrum fixed.
}
    \label{tab:cutflow_had}
\end{table}

Finally, we briefly consider the effect of hadronization uncertainties. The aim of this comment is not to present a full account of hadronization uncertainties but rather provide an indication of the cuts most susceptible to it. In our setup, $\ld$ serves as a proxy for overall confinement scale as well as the scale at which one loop running coupling diverges. However, such a definition is not exact. As $\ld$ also controls hadronization cut-off, any variation of this definition would change the hadron multiplicity and thus presents itself as a source of hadronization uncertainty. We therefore vary $\ld$ by a factor of 3 while keeping the physical observables i.e. $\pi$ and $\rho$ masses fixed. The resulting variation in the cutflow efficiencies is shown in tab.~\ref{tab:cutflow_had}. 

We observe that the variation in $\ld$ affects efficiencies non-trivially. The most significant variation is observed for 1 trk, $|d_0|>3$mm while the next relevant cut is $m_{DV}/\Delta R>4$ GeV. This illustrates the importance of evaluating hadronization uncertainties specially for analyses relying on track and vertex features. However we also note here that the final efficiencies of our analysis do not vary by more than 20\% for the $\ld$ variation considered here. For a more comprehensive treatment of hadronization uncertainties see~\cite{Liu:2025bbc}.

\section{Results}
\label{sec:results}
We now present our results for the proposed SVEJ analysis along with comparison with two existing analyses, the ATLAS CalRatio (CR)~\cite{ATLAS:2024ocv} and the CMS muon detector shower (MDS)~\cite{CMS:2021juv} analysis. The reinterpretation of CMS muon detector shower has been previously discussed in detail in ~\cite{Liu:2025bbc}, therefore the discussions here are brief and shown for completeness. 

The dark pion lifetime of our interest may potentially be probed via several existing LHC analyses even if they were not designed to search for dark showers. Somewhat ironically, as we stated in section~\ref{sec:gen_level}, the signal of our interest is not efficiently being probed by the current emerging jet analyses as they either require hard objects in the final state or rely on missing energy triggers, which we show to be less efficient than the emerging jet trigger in table~\ref{tab:triggers}. 

Along with the emerging jets analyses, generic searches for long-lived particles may be of an interest for this model. For example, a number of searches aim to detect LLPs in the inner part of the detector~\cite{ATLAS:2023oti, CMS:2021tkn, CMS:2024trg,CMS:2020iwv}. Many of these searches are optimised for a susy-like signal with a heavy LLP in mind and thus trigger on large MET or $H_T$. These triggers often have much higher threshold and lead to low acceptance for our signal. Among these ~\cite{CMS:2021tkn} employs a similar strategy as our proposal, albeit for the case of two LLPs and thus assumes that the displaced vertices have large angular separation. As stated before we use~\cite{CMS:2021tkn} as our inspiration. The other class of searches target $H \to XX$ topologies among which~\cite{ATLAS:2024qoo, ATLAS:2021jig,CMS:2021yhb} target SM Higgs production in association with a SM gauge boson, and hence are not applicable to the signal under consideration here. Finally~\cite{CMS:2024xzb} could be relevant, and we do consider its trigger strategy to understand its importance for our signal topology.

Beyond the searches in the inner detector, in recent years, a number of innovative searches have been carried out for hadronically decaying LLPs in the outer parts of the detector. Among them, the CMS MDS search (long-lived particles decaying in the muon system)~\cite{CMS:2021juv} was reinterpreted in~\cite{Liu:2025bbc}, for the same model but with longer lifetimes. In a similar spirit falls the ATLAS CR search, which targets hadronically decaying long-lived particles in association with jets or leptons by exploiting the ratio of the energy deposits in the hadronic calorimeter to the energy in the electromagnetic calorimeter, using the full Run-2 dataset~\cite{ATLAS:2024ocv}. We discuss in the next subsection the detailed reinterpretation of these two searches.

\subsection{Reinterpretation strategies for CalRatio and muon displaced shower analyses}
The CMS analysis we use here is documented in~\cite{CMS:2021juv},\footnote{An update of the search is available in~\cite{CMS:2024bvl} however the associated reinterpretation material does not include a {\tt Delphes} code or module implementation, hence we use the older version of the analysis.}, with the associated reinterpretation material including a {\tt Delphes} implementation provided in~\cite{CMSreinterpretation}. The analysis searches for decays of long-lived particles (LLPs) in the CMS muon system, in particular the muon endcap detector. The analysis is optimized for single or pair production of LLPs. Any LLP decaying in the CMS muon system, specifically in the CMS muon endcap detector, induces hadronic and electromagnetic showers, giving rise to high hit multiplicity in the localized regions of the detector called the CSC clusters. In particular, a CSC cluster associated with the signal requires $N_{\rm hits} > 130$ and the azimuthal angle between the cluster location and the MET ($\Delta \phi_c$) $ < $ 0.75.  The analysis searches for these CSC clusters by using the muon endcap detector as a sampling calorimeter. The search is sensitive to LLPs decaying to hadrons, taus, electrons, or photons. Decays to muons are not considered as muons rarely produce particle shower which lead to the CSC cluster. 

The search requires ${\rm MET} > 200~\rm{GeV}$ (where MET is defined as the negative vector sum of visible $p_T$ from particles identified in the tracker and calorimeter), no electron~(muon) with transverse momentum $p_T > 35~(25)$~GeV and pseudo-rapidity $|\eta| < 2.5~(2.4)$, at least one CSC cluster with $|\Delta\phi_c| < 0.75$ and no muons or electrons close to the clusters.  Finally, since the CSC clusters are used as calorimeters, the cluster efficiency depends on the amount of electromagnetic and hadronic energy deposited in the muon system. The search therefore provides cluster efficiency as a function of the hadronic and electromagnetic energy deposited in the Muon System.

In order to reinterpret this analysis we pass our events through \textsc{Delphes}~v3.5.1 \cite{deFavereau:2013fsa} and modify the associated \textsc{Delphes} card as described in~\cite{Liu:2025bbc}.

The original CalRatio search targets long-lived neutral scalars in the Hidden Abelian Higgs model. The analysis is capable of searching for either one or two LLPs. Nevertheless, a re-interpretation tool is provided, which can recast the selection efficiency from the original interpretation to any other model, using the machine-learning paradigm of surrogate modeling~\cite{HEPData, Corpe:2025sbw}. They are implemented as multi-class Boosted Decision Trees. The tool takes truth-level LLP decay position and kinematics for the leading or leading and sub-leading LLPs, as well as the particle ID of the leading decay products as an input. This information can be obtained following the common setups of simulation using \texttt{MadGraph} and \texttt{PYTHIA}. Subsequently, for models such as the dark shower scenario considered here, the tool outputs the selection efficiency of multiple final states in various regions of the ATLAS detector. Among them, we take the output of ``CalRatio + 2J" final state, which means that the displaced jet is associated with an unusual calorimeter energy ratio (CalRatio) and is accompanied by two resolved jets (2J). This choice is made since this final state is most relevant to our dark shower models. The other final states are CalRatio + W/Z bosons, which additionally requires the kinematics of the gauge bosons as input and hence are not suitable for our model. Following the above procedures, we obtain the efficiency, as well as the subsequent upper limits on the cross section, for the dark shower model, from the ATLAS CR searches.

\subsection{Limits as a function of $\ctp$}
\begin{figure}[h!]
    \centering 
    \includegraphics[width=0.49\linewidth]{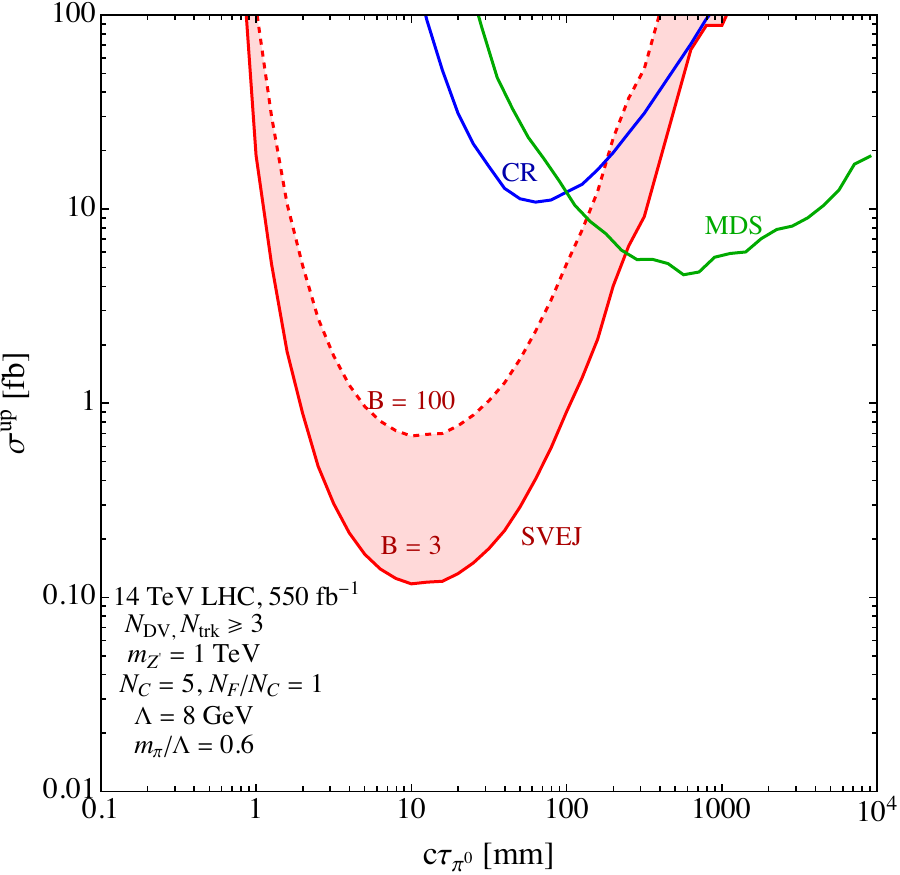}
    \includegraphics[width=0.49\linewidth]{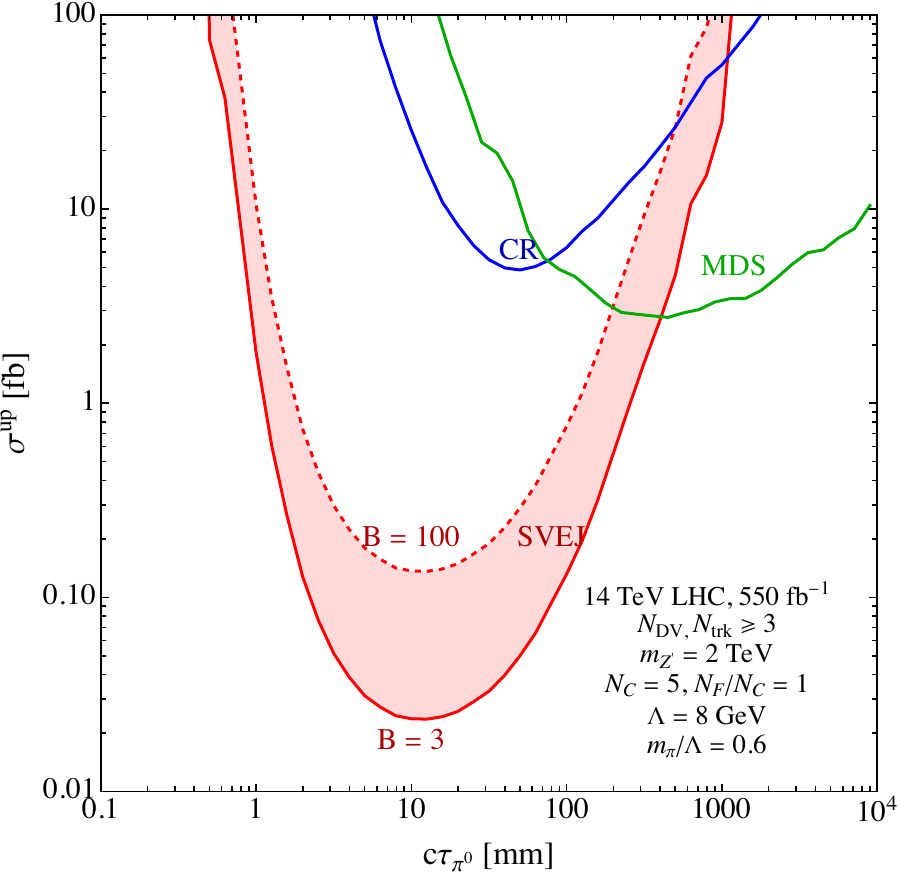}
    \caption{Comparison of upper limits on the signal cross-section obtained using CR analysis and the proposed SEVJ analysis as a function of the dark pion lifetime $\ctp$ for $\mzp = 1$ TeV (left panel) and 2 TeV (right panel) for fixed $\nc = 5, \fc = 1, \mpl = 0.6, \ld = 8\,\rm{GeV}$. For the SEVJ analysis, we also show the projection derived assuming 3 and 100 background events. Upper limits are shown for signal region $N_{trk} \geq 3, N_{DV} \geq 3$.}
    \label{fig:runct}
\end{figure}

In fig.~\ref{fig:runct}, we first show comparison of the upper limits for the three different analyses as a function of the LLP lifetime $\ctp$. Along with the CalRatio (CR) and the semi-visible emerging jets (SVEJ), we also include the projected upper limits for the CMS muon displaced shower (MDS). The behavior of MDS analysis was studied in detail in ~\cite{Liu:2025bbc}, we therefore do not spend more time on it here. As we can not estimate the backgrounds for the proposed emerging SVJ analysis, we show a band encompassed by upper limits assuming 3 and 100 background events. As expected our proposed analysis probes smaller lifetimes compared to the CR analysis. It should be kept in mind that the CR analysis was not designed for the signal under consideration, therefore the effects of having multiple long lived particles depositing energies in different calorimeter regions are as yet unclear. For our purposes, we have ignored this complication. Comparing the upper limits for two $Z^\prime$ masses, we observe that the upper limits get stronger for heavier $Z^\prime$ for all three analyses. This is expected due to increased multiplicity of the dark pions. The proposed emerging SVJ analysis exhibits a larger change in the upper limits for the two $Z^\prime$ masses compared to the CR or MDS analysis. This stems from the relative change in the efficiency where the emerging-jet trigger we use for the SVEJ analysis gets more efficient at larger $Z^\prime$ masses.

\begin{figure}[h!]
    \centering 
    \includegraphics[width=0.49\linewidth]{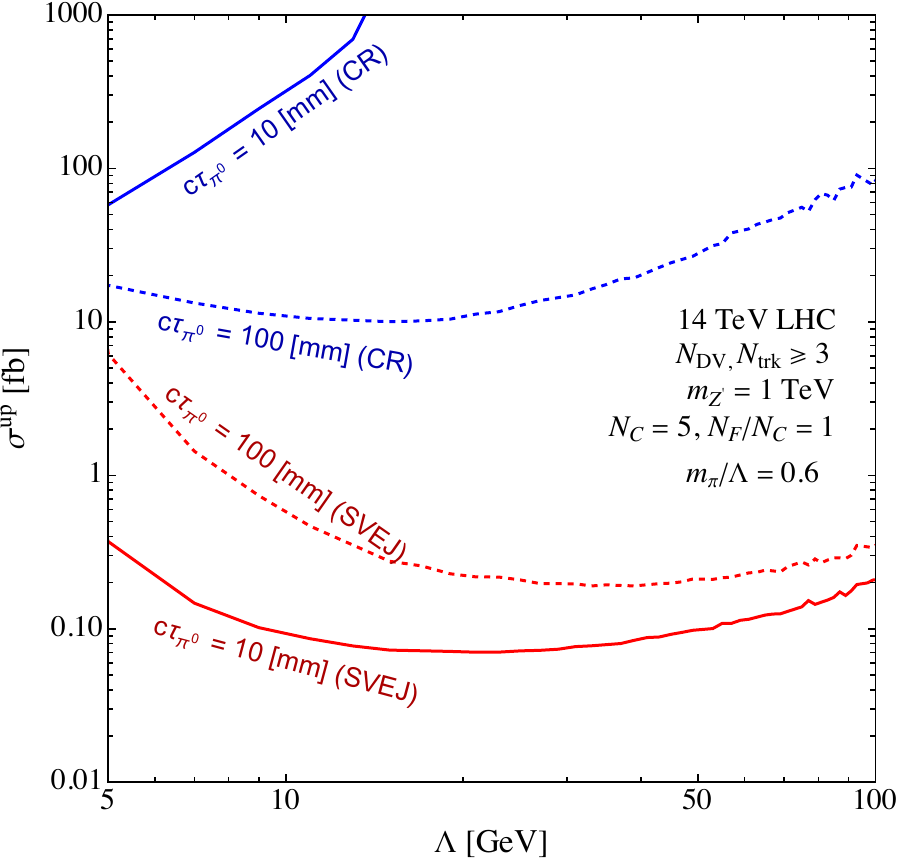}
    \includegraphics[width=0.49\linewidth]{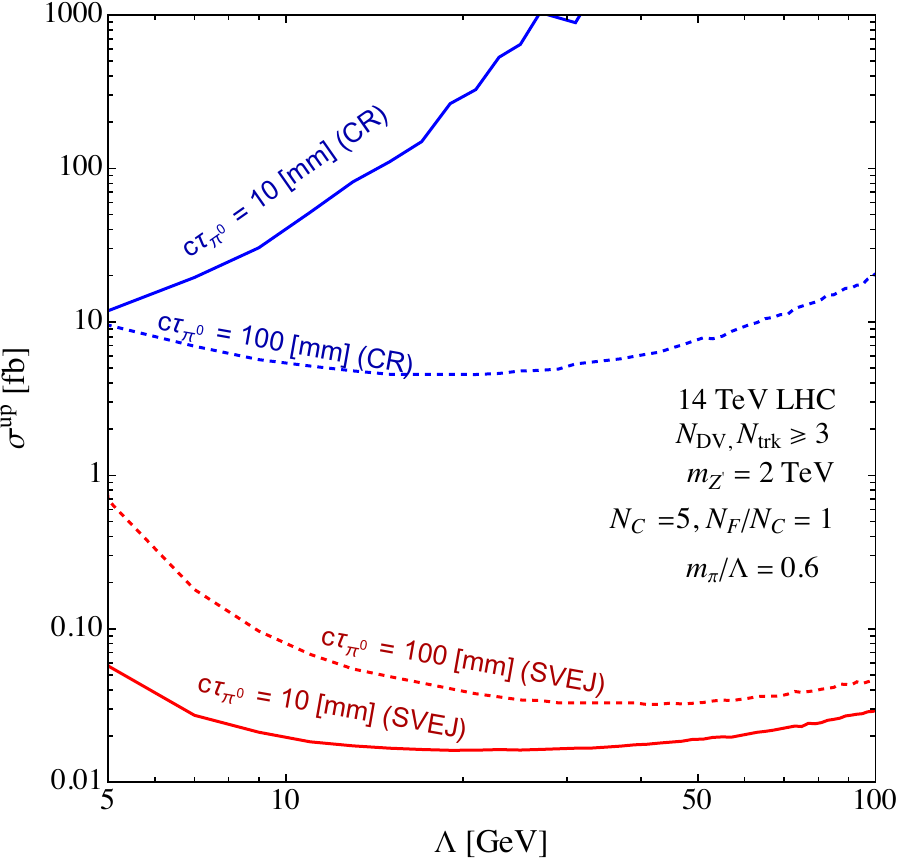}
    \caption{Comparison of upper-limits obtained for CR and SVEJ analyses for $\mzp = 1$ TeV (left panel) and $\mzp = 2$ TeV (right panel) for $\ctp = 10$mm (solid lines) and 100 mm (dashed lines). The bands represent different assumptions on number of signal events applicable for SVEJ analysis. }
    \label{fig:runlambda}
\end{figure}

It is also interesting to note that the upper limits obtained by the SVEJ analysis are the strongest among the three, followed by the MDS and then the CR analysis. This can be understood by comparing the relative background events as well as signal efficiencies. For the CR analysis the scaled backgrounds corresponds to 400 background events,  while for the MDS analysis there are 8 background events. Therefore, compared to the SVEJ analysis, the CR has larger backgrounds and correspondingly worse limits. If we assume 400 background events for the SVEJ analysis, the difference in the CR and SVEJ limits corresponds to selection efficiencies. Despite the low background, the MDS analysis has poor selection efficiency since our events lack the required high-energy muon deposits, and thus it does not outperform the proposed SVEJ limits. We do not consider MDS analysis for the discussion below since a dedicated discussion is already available elsewhere~\cite{Liu:2025bbc}. 

\begin{figure}[h!]
    \centering 
    \includegraphics[width=0.49\linewidth]{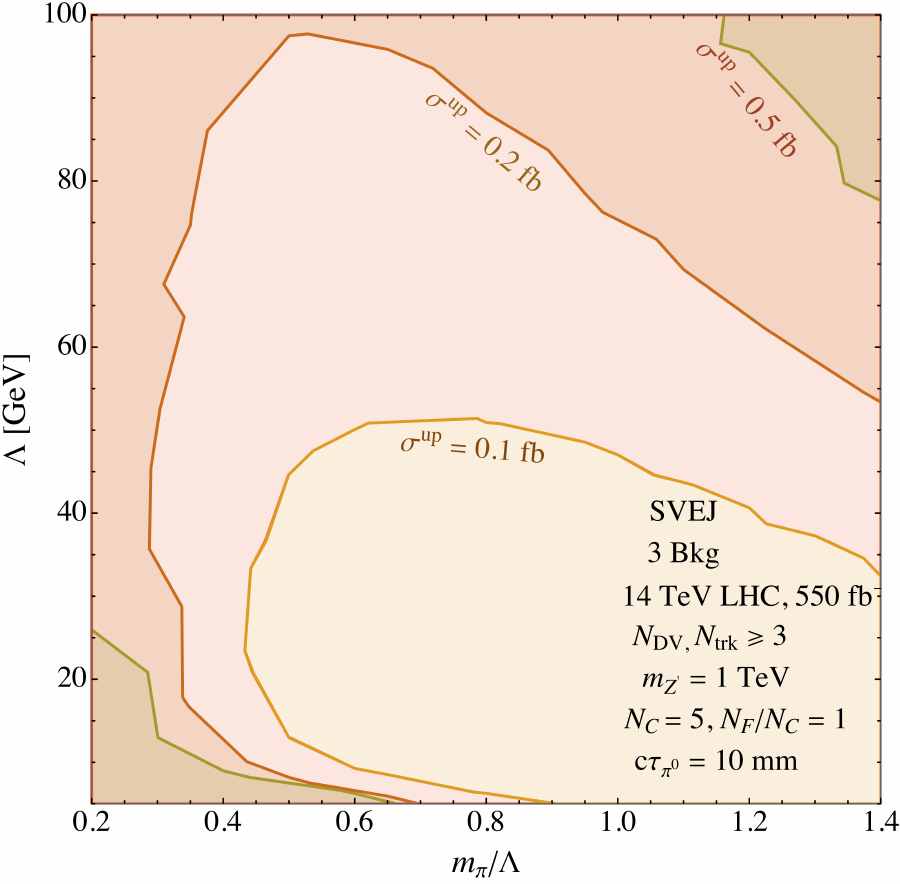}
    \includegraphics[width=0.49\linewidth]{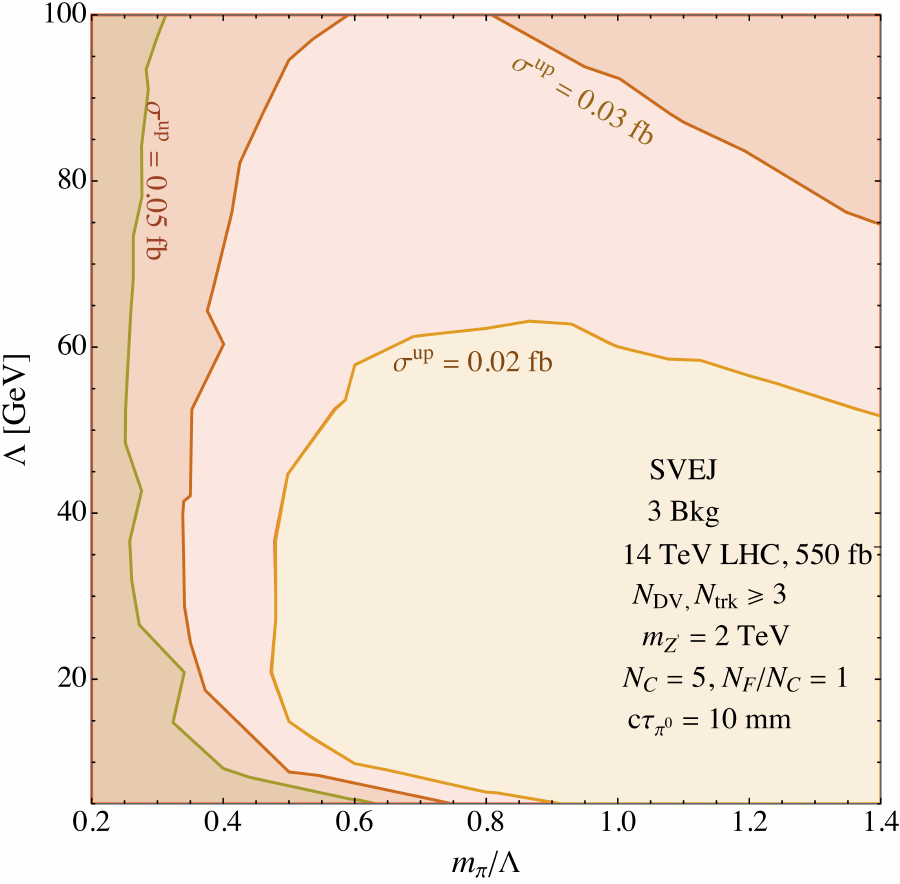}
    \caption{Upper limits as a function of $\mpl$ and $\ld$ obtained using SVEJ analysis for fixed $\ctp = 10$ mm, and $\mzp = 1$ TeV, (left panel), and $\mzp = 2$ TeV (right panel).}
    \label{fig:2D_upper_limits_ESVJ}
\end{figure}
Fig.~\ref{fig:runlambda}, shows comparison of upper limits obtained with the CR and SVEJ analysis as a function of the overall scale of the theory $\ld$ for two different $Z^\prime$ masses and two different lifetimes. Concentrating on the case of $\mzp = 1$ TeV, we observe that CR analysis has weaker upper limits for smaller $\ctp$. This is expected given results in fig.~\ref{fig:runct}. At $\ctp = 10$ mm, the CR analysis yields $\sigma^{up} < 1000$ fb for $\ld < 20$ GeV. This is also expected as the pion multiplicity and boost is large for such small $\ld$ and thus some of the pions may reach calorimeter, depositing energy there. Once the lifetime increases to 100 mm, many pions may reach calorimeter and thus the analysis leads to $\sigma^{up} < 10$ fb over a broad range of $\ld$. On the contrary, with the proposed SVEJ analysis, we obtain stronger upper limits across all values of $\ld$ for both the lifetimes. Increasing the $\mzp$ to 2 TeV, we broadly see similar trends, although generically we obtain stronger upper limits compared to the 1 TeV scenario. This is primarily because of the increased boost. Finally, we make an important observation about the behavior of upper limits for SVEJ analysis. The upper-limits follow a shallow parabola as a function of $\ld$. For a fixed $\mzp$, if $\ctp$ increases, the parabolic shape remains the same but the minima shifts to higher $\ld$. Although not shown here, a similar behavior is seen in $\mpl$ plane. 

\subsection{Limits as a function of $\ld - \mpl$}
We now turn our attention to the behavior of upper limits in $\mpl - \ld$ plane for the two different analyses. In fig.~\ref{fig:2D_upper_limits_ESVJ} we show the projected upper limits for upcoming LHC Run - 4 with integrated luminosity $\mathcal{L} = 550\,\rm{fb}^{-1}$, see~\cite{Tomas:2022mzg} for details.  The projections assume 3 background events. A notable feature of the upper-limit behavior is the presence of closed contours, suggesting degeneracies in the parameter space where identical upper limits occur for different values of $\ld$ and $\mpl$. This is however not surprising given the results in fig.~\ref{fig:runlambda}, which clearly demonstrate that it is possible to obtain same upper limits for two different values of $\ld$. Increasing the value of $\mzp$ to 2 TeV leads to an order of magnitude better sensitivity in upper limits, mirroring the discussions in ref.~\ref{fig:runct} and ~\ref{fig:runlambda}. 

\begin{figure}[h!]
    \centering 
    \includegraphics[width=0.49\linewidth]{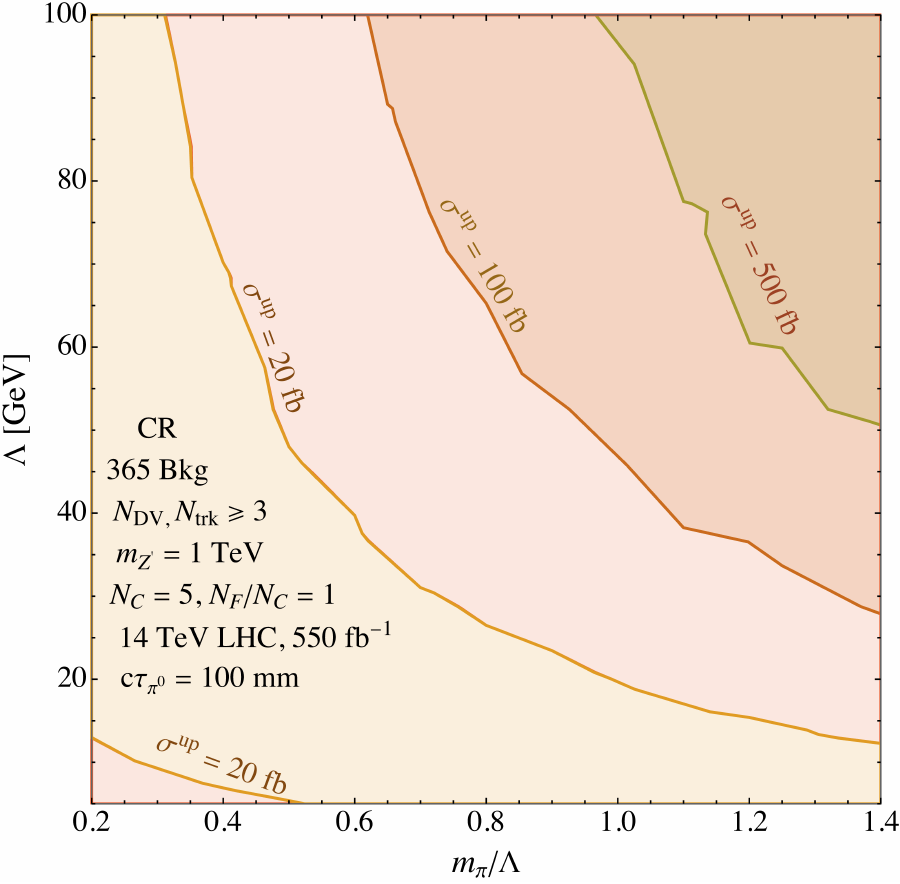}
    \includegraphics[width=0.49\linewidth]{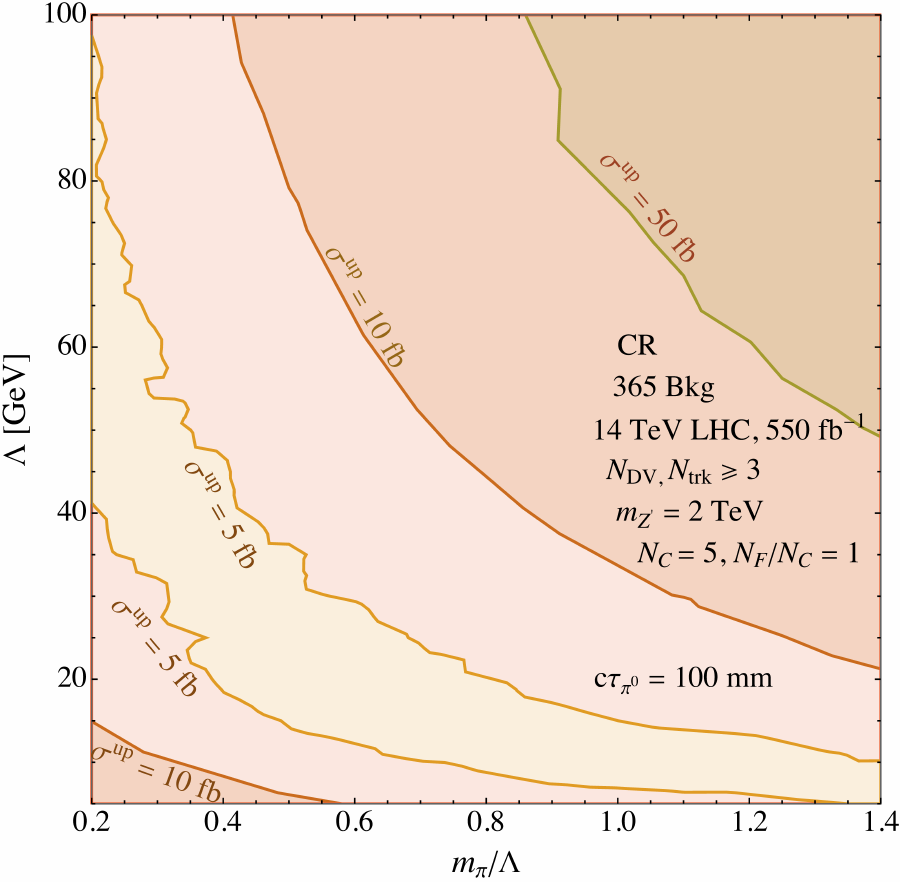}
    \caption{Upper limits as a function of $\mpl$ and $\ld$ obtained using CR analysis for fixed $\ctp = 100$ mm, and $\mzp = 1$ TeV, (left panel), and $\mzp = 2$ TeV (right panel). }
    \label{fig:2D_upper_limits_CR}
\end{figure}

In fig.~\ref{fig:2D_upper_limits_CR}, we show the upper limits obtained using the CR analysis for fixed $\ctp = 100$ mm, and $\mzp = 1$ TeV, (left panel), and $\mzp = 2$ TeV (right panel). $\ctp = 100$ mm corresponds to the lifetime at which the analysis gets strongest upper limits. This plot assumes 365 background events, obtained by using luminosity scaling. We note here that the upper limits here are not comparable with fig.~\ref{fig:2D_upper_limits_ESVJ} because the two projections are obtained for different $\ctp$~\footnote{Using a common $\ctp$ requires generation of a very large number of events without necessarily adding extra information.}. The upper limits behavior here is reminiscent of~\cite{Liu:2025bbc}. The most striking difference between the results presented in ~\cite{Liu:2025bbc} and this result is again the appearance of degenerate upper limits which is more prominently visible at higher $\mzp$. This behavior is once again explained by results in fig.~\ref{fig:runlambda}, where parabolic shape for upper limit is obtained for larger $\ctp$. It is curious to note that such two sided behavior was not seen in ref~\cite{Liu:2025bbc}. In that particular analysis, the strongest upper limits were obtained for $\ctp \sim \mathcal{O} (500)\,\rm{mm}$, however the equivalent 2D plots were made for $\ctp = 100\,\rm{mm}$. As this is smaller than the point of maximal sensitivity, the second side of upper limits may be obtained for even smaller values of $\ld$ and hence are not visible on plots in ref.~\cite{Liu:2025bbc}. 

\section{Conclusions}
\label{sec:conclusions}

In anticipation of the upcoming LHC Run-3 and the renewed interest in dark shower phenomenology, a systematic classification of the accessible signature space, consistent with theoretical constraints, is essential for formulating a coherent experimental strategy. Guided by this motivation, we have proposed a new class of experimental signatures and an associated search strategy.

Focusing on the $s$-channel $Z^\prime$ production mechanism, we study a scenario in which the diagonal dark pions are unstable while the off-diagonal states, protected by dark-flavor symmetry, remain stable. Due to the small coupling between the Standard Model (SM) and the dark sector, the dark pions are naturally long-lived. Considering dark-pion decays occurring within the tracker, this interplay between stable and long-lived components gives rise to a little studied dark-shower signature, which we term semi-visible emerging jets (SVEJ).

We propose an analysis strategy targeting SVEJ signatures using the number of displaced vertices, associated tracks, and the invariant mass of displaced vertices as primary discriminating variables. Several trigger strategies are examined, including conventional missing transverse energy ($E_T^{\text{miss}}$) and $H_T$ triggers, as well as the displaced-jet and emerging-jet triggers proposed by ATLAS and CMS. Among these, the ATLAS emerging-jet trigger is found to be best suited to our analysis, though further optimization could enhance future sensitivity.

Our results indicate that the proposed analysis is sensitive to dark-sector couplings $\ld < 10~\text{GeV}$, where multiple dark-pion production becomes significant, and to dark-pion lifetimes of order $10~\text{mm}$.

Additionally, we reinterpret the existing ATLAS CalRatio analysis, which targets decays of long-lived particles in the calorimeter, corresponding to dark-pion lifetimes of order $100~\text{mm}$. Although this analysis yields weaker upper limits than our proposed SVEJ search, it complements our results by bridging the gap between our sensitivity region and that probed by the CMS muon-shower analysis.

Together, these studies highlight the importance of exploring DS parameter space systematically, motivating dedicated searches in the upcoming LHC data-taking runs.

\section*{Acknowledgments}
We would like to thank Lisa Benato, Jackson Burzynski,  Bryan Cardwell, Louie Dartmoor Corpe, Alberto Escalante del Valle, Ali Kaan G\"uven, Ang Li, Emma Torr\'o and Zhenyu Wu for useful discussions, and  Lisa Benato, Jackson Burzynski, and Louie Dartmoor Corpe for a careful reading of the manuscript.  SK and JL are supported by the FWF project number P 36947-N.
JC and JZ are supported by the {\it Generalitat Valenciana} (Spain) through the {\it plan GenT} program (CIDEGENT/2019/068), by the Spanish Government (Agencia Estatal de
Investigaci\'on), ERDF funds from European Commission (MCIN/AEI/10.13039/501100011033, Grant No. PID2020-114473GB-I00), and by the Spanish Research Agency (Agencia Estatal
de Investigaci\'on, MCIU/AEI) through the grant IFIC Centro de Excelencia Severo Ochoa No. CEX2023-001292-S. SK, JZ and JC also acknowledge the support from the Consejo Superior de Investigaciones Cient\'ificas (CSIC) through the ``Red de Internacionalizaci\'on'' ILINK, grant No. 22043, which enabled this project. WL is supported by National Natural Science foundation of China (Grant
No. 12205153).
JC is also supported by CIBEFP/2023/120, and is very grateful for the hospitality of the theoretical particle physics group at the University of Graz, where the initial phase of this project was carried out.

\bibliography{bibliography}

\end{document}